\documentclass[twocolumn,showpacs,amsmath,amssymb]{revtex4}
\usepackage{graphicx,url,hyperref,dcolumn,bm,lineno}
\hypersetup{colorlinks,citecolor=blue,linkcolor=red,urlcolor=blue}
\begin{document}
%--------------------------------------------------------------------
\title{Flow-Area Relations in Immiscible Two-Phase Flow in Porous Media}
%--------------------------------------------------------------------

\author{Subhadeep Roy${}^{1}$}
\email{subhadeep.roy@ntnu.no}
\author{Santanu Sinha${}^2$}
\email{santanu@csrc.ac.cn}
\author{Alex Hansen${}^{1,2}$}
\email{alex.hansen@ntnu.no}
\affiliation{
${}^1$PoreLab, Department of Physics, Norwegian University of 
Science and Technology, NO--7491 Trondheim, Norway. \\
${}^2$Beijing Computational Science Research Center, 
10 East Xibeiwang Road, Haidian District, Beijing 100193, China.}

%--------------------------------------------------------------------
\date{\today {}}
%--------------------------------------------------------------------
\begin{abstract}
We present a theoretical framework for immiscible incompressible
two-phase flow in homogeneous porous media that connects the
distribution of local fluid velocities to the average seepage 
velocities.  By dividing the pore area along a cross-section transversal to
the average flow direction up into differential areas associated 
with the local flow velocities, we construct a distribution function
that allows us not only to re-establish existing relationships 
between the seepage velocities of the immiscible fluids, but also
to find new relations between their higher moments. We support 
and demonstrate the formalism through
numerical simulations using a dynamic pore-network model for
immiscible two-phase flow with two- and three-dimensional pore networks. 
Our numerical results are in agreement with the theoretical considerations.
\end{abstract}
%--------------------------------------------------------------------
\maketitle
%--------------------------------------------------------------------

%--------------------------------------------------------------------
\section{Introduction}
\label{intro}
When two immiscible fluids compete for the same pore space, we are 
dealing with immiscible two-phase flow in porous media \cite{b88}.    
A holy grail in porous media research is to find a proper description of
immiscible two-phase flow at the continuum level, i.e., at scales where
the porous medium may be treated as a continuum.  Our understanding of 
immiscible two-phase flow at the pore level is increasing at a very high 
rate due to advances in experimental techniques combined with an explosive 
growth in computer power \cite{b17}.  Still, the gap in scales between
the physics at the pore level and a continuum description remains huge and
the bridges that have been built so far across this gap is either complex to
cross or rather rickety.  To the latter class, we find the still dominating 
theory, first proposed by Wyckoff and Botset in 1936 \cite{wb36} and with an 
essential amendment by Leverett in 1940 \cite{l40}, namely relative permeability 
theory.  The basic idea behind this theory is the following: Put yourself in the
place of one of the two immiscible fluids.  What does this fluid see?  It sees
a space in which it can flow limited by the solid matrix of the porous medium, 
but {\it also by the other fluid.\/}  This reduces its mobility in the porous
medium by a factor known as the relative permeability, which is a function of 
how much space there is left for it.  And here is the rickety part: this
reduction of available space --- expressed through the saturation --- is the
{\it only\/} parameter affecting the reduction factor or relative permeability. 
This is a very strong statement and clearly does not take into account that
the distribution of immiscible fluid clusters will depend on how fast the fluids
are flowing.  Still, in the range of flow rates relevant for many industrial 
applications, this assumption works pretty well. It therefore remains the essential 
work horse for practical applications.

Thermodynamically Constrained Averaging Theory (TCAT) \cite{hg90,hg93,hg93b,nbh11,gm14} is
a built on the framework of relative permeability. However, it is based on a full 
analysis based on mechanical conservation laws, constitutive laws, e.g.\ for 
the motion of interfaces and contact lines, and on thermodynamics at the pore 
level.  These are then scaled up using averaging theorems, which, loosely 
explained, consist of replacing derivatives of averages by averages of 
derivatives.  In principle, this approach solves the up-scaling problem.  However, 
as Gray and Miller point out in their book \cite{gm14}, each component of TCAT 
involves significant mathematical manipulations.  The internal energy has 
contributions from the bulk liquids, the fluid-fluid and fluid-matrix interfaces 
at the pore level.  The averaging process redefines the variables describing 
these contributions, but does not reduce their number.  This accounts for a 
high level of complexity.  

A further development somewhat along the same lines, based on non-equilibrium 
thermodynamics uses Euler homogeneity, more about this later, to define 
the up-scaled pressure.  From this, Kjelstrup et al.\ derive constitutive 
equations for the flow \cite{kbhhg19,kbhhg19b}.

Another class of theories is based on detailed and specific assumptions 
concerning the physics involved. An example is Local Porosity Theory
\cite{hb00,h06a,h06b,h06c,hd10,dhh12}.  Another is DeProf theory which is 
a mechanical model combined with non-equilibrium statistical 
mechanics based on a classification scheme of fluid configurations 
at the pore level \cite{vcp98,v12,v18}.

A recent work \cite{hsbkgv18} has explored a new approach to immiscible 
two-phase flow in porous media based on elements borrowed from thermodynamics.
That is, it is using the framework of thermodynamics, but without 
connecting it to processes involving heat.  The spirit behind this 
approach is similar to that of taken in Edwards and Oakshot's 
pseudo-thermodynamic theory of powders \cite{eo89}.  The approach
consists in looking for general relations that transcends details of the
physical processes involved.  An example of such an approach in 
the field of immiscible two-phase flow in porous media is found in
the Buckley-Leverett theory of invasion fronts \cite{bl42}.  The
Buckley-Leverett equation is based solely on the principle of mass conservation
and on the fractional flow rate being a function of the saturation. In the
approach of Hansen et al.\ \cite{hsbkgv18}, equations are derived that
originate from Euler homogeneity as in ordinary thermodynamics.  These
equations transcend the details of the physics involved in the same way
that the equations of thermodynamics are universally applicable as long as
a set of simple underlying conditions are met. 

Thermodynamics is a theory that is valid on scales large enough so that
the system it refers to may be regarded as a continuum.  Statistical mechanics 
is then the theory that makes the connection between thermodynamics and
the underlying atomistic picture.

It is the aim of this paper to formulate a description of immiscible two-phase
flow in porous media that may form a link between the continuum-level approach
of Hansen et al.\ \cite{hsbkgv18} and the pore-level description of the problem
--- a sort of ``statistical mechanics" from which the pseudo-thermodynamics
may be derived, but which also describes the flow problem at the pore level. 
  
After defining the system and the variables involved in Section \ref{secDef},
we will in Section \ref{secThermo} review the pseudo-thermodynamic approach
\cite{hsbkgv18}. The next Section \ref{secDiff} we introduce the central
object in the paper, the {\it differential transversal area distribution\/}
which corresponds to the Boltzmann distribution in ordinary statistical
mechanics, and relate it to the pseudo-thermodynamics relations.  Then
follows Section \ref{secMoments} which then moves beyond the 
results of the pseudo-thermodynamics by focusing on fluctuations.  In
Section \ref{secNum} we use the dynamic network simulator \cite{jh12}
first introduced by Aker et al.\ \cite{amhb98} and then later refined
\cite{gvkh19,gwh19,sgvh19} to verify the relations derived in the 
earlier sections.  There is also a second goal behind this numerical
work: the dynamic network model is a model at pore level and by its use, 
we show how the formalism developed here connect to the flow patterns
at the pore level. Finally, we draw our conclusions in Section \ref{secCon}.
%--------------------------------------------------------------------
\section{System definition}
\label{secDef}

In two-phase flow, the steady state \cite{tkrlmtf09,tlkrfm09,aetfhm14} is
characterized by potentially strong fluctuations at the pore scale,
but steady averages at the REV (Representative Elementary Volume) 
scale. As such they differ
fundamentally from stationary states that are static at the pore scale
as well. Steady states have much in common with ensembles in
equilibrium statistical mechanics. They are also by implication
assumed in the conventional descriptions of porous media flows that
take the existence of an REV for granted.

Our REV is a block of homogeneous porous material of length $L$ and
area $A$. We prevent flow through the surfaces that are parallel to
the $L$-direction which is the flow direction. The two remaining
surfaces, each having an area $A$, act as inlet and outlet for the
incompressible fluids that are injected and extracted from the
REV. The porosity of the material is defined as
\begin{equation}
\label{eqn0}
\phi = \frac{V_p}{V}\;,
\end{equation}
where $V_{p}$ is the pore volume and $V=AL$ is the volume of the REV. 
Due to the homogeneity of the porous medium, any cross section orthogonal 
to the axis along the $L$-direction will have a pore area that
fluctuates around the value 
\begin{equation}
\label{eqn0.01}
A_{p}=\frac{V_p}{L}=\phi A\;.
\end{equation} 
There is also a solid matrix area fluctuating around 
\begin{equation}
\label{eqn0.25}
A_s=A-A_{p}=(1-\phi )A\;. 
\end{equation}
The homogeneity assumption consists in the fluctuations being so small
that they can be ignored.

There is a time averaged volumetric flow rate $Q$ through the REV.
The volumetric flow rate consists of two components, $Q_{w}$ and
$Q_{n}$, which are the volumetric flow rates of the more wetting ($w$
for ``wetting") and the less wetting ($n$ for ``non-wetting") fluids
with respect to the porous medium. They are related through
\begin{equation}
\label{eqn1}
Q=Q_{w}+Q_{n}\;.  
\end{equation}

In the porous medium, there is a volume $V_w$ of the wetting fluid and
a volume $V_n$ of the non-wetting fluid so that $V_p=V_w+V_n$. We
define the wetting and non-wetting saturations $S_w=V_w/V_p$ and
$S_n=V_n/V_p$, so that $S_w+S_n=1$.

We define the wetting and non-wetting transversal pore areas $A_{w}$ and $A_{n}$
as the parts of the transversal pore area $A_{p}$ which occupied by the wetting or
the non-wetting fluids respectively. We have that 
\begin{equation}
\label{eqn5}
A_{p}=A_{w}+A_{n}\;.  
\end{equation}
As the porous medium is
homogeneous, we will find the same averages $A_{w}$ and $A_{n}$ in any
cross section through the porous medium orthogonal to the flow
direction. We have therefore
$A_{w}/A_{p}=(A_{w}L)/(A_{p}L)=V_{w}/V_{p}=S_{w}$, so that
\begin{equation}
\label{eqn3}
A_{w}=S_{w}A_{p}\;.  
\end{equation}
Likewise,
\begin{equation}
\label{eqn4}
A_{n}=S_{n}A_{p}=\left(1-S_{w}\right) A_{p}\;. 
\end{equation}

We define the seepage velocities, i.e., the average flow velocities
in the pores, for the two immiscible fluids, $v_w$
and $v_n$ as
\begin{equation}  
\label{eqn5.1}
v_w=\frac{Q_w}{A_w}\;,
\end{equation}
and 
\begin{equation}  
\label{eqn5.2}
v_n=\frac{Q_n}{A_n}\;.
\end{equation}
The seepage velocity associated with the total flow rate $Q$ is
defined as
\begin{equation}  
\label{eqn5.3}
v_p=\frac{Q}{A_p}\;.
\end{equation}
We may express equation (\ref{eqn1}) in terms of the seepage 
velocities,
\begin{equation}  
\label{eqn5.4}
v_p=S_w v_w+S_nv_n\;.
\end{equation}

%--------------------------------------------------------------------
\section{Pseudo-Thermodynamic relations}
\label{secThermo}

Hansen et al.\ \cite{hsbkgv18} derived a number of relations between
the seepage velocities defined in (\ref{eqn5.1})--(\ref{eqn5.3})
based on the volumetric flow rate being
an Euler homogeneous function of order one with respect to the wetting
and non-wetting transversal pore areas $A_w$ and $A_n$, meaning that the volumetric flow
rate obeys the scaling law
\begin{equation}
\label{eqn0.100}
Q(\lambda A_w,\lambda A_n)=\lambda Q(A_w,A_n)\;,
\end{equation}
where $\lambda$ is a scale factor. By taking the derivative of this equation 
with respect to $\lambda$ and then setting $\lambda=1$, we find
\begin{equation}
\label{eqn0.101}
Q(A_w,A_n)=\left(\frac{\partial Q}{\partial A_w}\right)_{A_n}A_w+
\left(\frac{\partial Q}{\partial A_n}\right)_{A_w}A_n\;.
\end{equation}
By dividing this expression by the transversal pore area $A_p$ and using
equations (\ref{eqn5}) to (\ref{eqn4}), we may write this equation as
\begin{equation}
\label{eqn0.3}
v_p= S_w\left(\frac{\partial Q}{\partial A_w}\right)_{A_n}A_w+
S_w\left(\frac{\partial Q}{\partial A_n}\right)_{A_w} \;.
\end{equation}
The two partial derivatives have the units of velocity, and 
Hansen et al.\ \cite{hsbkgv18} name these velocity functions
the {\it thermodynamic velocities,\/}
\begin{equation}
\label{eqn0.1}
\hat{v}_{w}=\left(\frac{\partial Q}{\partial A_{w}}\right)_{A_{n}}\;,  
\end{equation}
and
\begin{equation}
\label{eqn0.2}
\hat{v}_{n}=\left(\frac{\partial Q}{\partial A_{n}}\right)_{A_{w}}\;.  
\end{equation}

We use equations (\ref{eqn3}) and (\ref{eqn4}) and the chain rule to derive
\begin{eqnarray}
\label{eqn0.1000}
\left(\frac{\partial}{\partial A_w}\right)_{A_n}
&=& \left(\frac{\partial S_w}{\partial A_w}\right)_{A_n}
\left(\frac{\partial }{\partial S_w}\right)_{A_p}\nonumber\\
&+&\left(\frac{\partial A_p}{\partial A_w}\right)_{A_n}
\left(\frac{\partial }{\partial A_p}\right)_{S_w}\nonumber\\
&=&\frac{S_n}{A_p}\ \left(\frac{\partial }{\partial S_w}\right)_{A_p}
+\left(\frac{\partial }{\partial A_p}\right)_{S_w}\;.
\end{eqnarray}
Likewise, we find 
\begin{equation}
\label{eqn0.1001}
\left(\frac{\partial}{\partial A_n}\right)_{A_w}
=-\frac{S_w}{A_p}\ \left(\frac{\partial }{\partial S_w}\right)_{A_p}
+\left(\frac{\partial }{\partial A_p}\right)_{S_w}\;.
\end{equation}
We now combine these two equations with the definitions (\ref{eqn0.1}) 
and (\ref{eqn0.2}), and use that $Q=A_pv_p$, i.e.\ equation (\ref{eqn5.3}), 
to find
\begin{equation}
\label{eqn0.3000}
\hat{v}_w=v_p+S_n\frac{dv_p}{dS_w}\;,
\end{equation}
and
\begin{equation}
\label{eqn0.3001}
\hat{v}_n=v_p-S_w\frac{dv_p}{dS_w}\;.
\end{equation} 

Combining the definitions (\ref{eqn0.1}) and (\ref{eqn0.2}) with equation (\ref{eqn0.3}) gives
\begin{equation}
\label{eqn5001}
v_p=S_w \hat{v}_w+S_n\hat{v}_n\;,
\end{equation}
which should be compared to equation (\ref{eqn5.4}). We see that
\begin{equation}
\label{eqn5002}
S_w v_w+S_n v_n =S_w \hat{v}_w+S_n\hat{v}_n\;.
\end{equation}

The seepage and
thermodynamic velocities are related through a transformation
$(v_w,v_n)\to(\hat{v}_w,\hat{v}_n)$ defining the {\it co-moving
velocity\/} $v_m$,
\begin{equation}
\label{eqn0.4}
\hat{v}_{w}=v_{w}+v_mS_{n}\;,  
\end{equation}
and 
\begin{equation}
\label{eqn0.5}
\hat{v}_{n}=v_{n}-v_mS_{w}\;.  
\end{equation}  

We now calculate
\begin{eqnarray}
\label{eqn0.50}
\left(\frac{\partial Q}{\partial S_w}\right)_{A_p}=
\left(\frac{\partial Q}{\partial A_w}\right)_{A_n}\left(\frac{\partial A_w}{\partial S_w}\right)_{A_p}
+\nonumber\\
\left(\frac{\partial Q}{\partial A_n}\right)_{A_w}\left(\frac{\partial A_n}{\partial S_w}\right)_{A_p}\;.
\end{eqnarray}
Using equations (\ref{eqn3}) and (\ref{eqn4}) together with equations (\ref{eqn0.3000}) and (\ref{eqn0.3001}),
we transform this equation into
\begin{equation}
\label{eqn18}
\frac{dv_p}{dS_{w}}=\hat{v}_{w}-\hat{v}_{n}\;,  
\end{equation}
where we have used that $v_p=Q/A_p$, i.e., equation ({\ref{eqn5.3}).  We now use equation 
(\ref{eqn5001}) to calculate
\begin{equation}
\label{eqn18.5}
\frac{dv_p}{dS_w}=\hat{v}_w-\hat{v}_n+S_w \frac{d\hat{v}_w}{dS_w}+S_n\frac{d\hat{v}_n}{dS_w}\;.
\end{equation}
Compare this equation to equation (\ref{eqn18}) and we get an analog to the Gibbs-Duhem equation,
\begin{equation}  
\label{eqn20}
S_w \frac{d\hat{v}_w}{dS_w}+S_n\frac{d\hat{v}_n}{dS_w}=0\;.
\end{equation}
Using equations (\ref{eqn0.4}) and (\ref{eqn0.5}), we find that the  
seepage velocities obey     
\begin{equation}
\label{eqn18-1}
\frac{dv_p}{dS_{w}}={v}_{w}-{v}_{n}+v_m\;,  
\end{equation}
and 
\begin{equation}  
\label{eqn20-1}
S_w \frac{d{v}_w}{dS_w}+S_n\frac{d{v}_n}{dS_w}=v_m\;.
\end{equation}
By combining equations (\ref{eqn0.1}), (\ref{eqn0.2}), (\ref{eqn0.4})
and (\ref{eqn0.5}), one finds
\begin{equation}
\label{eqn11-5.5}
v_w=v_p +S_n \left(\frac{dv_p}{dS_w}-v_m\right)\;,
\end{equation}
and 
\begin{equation}
\label{eqn11-5.6}
v_n=v_p -S_w \left(\frac{dv_p}{dS_w}-v_m\right)\;.
\end{equation}

These two equations, (\ref{eqn11-5.5}) and (\ref{eqn11-5.6}), may be seen
as a transformation $(v_p,v_m)\to (v_w,v_n)$.  The inverse of this
transformation, i.e., $(v_w,v_n)\to (v_p,v_m)$ are given by equations 
(\ref{eqn5.4}) and (\ref{eqn18-1}), i.e.,
\begin{eqnarray}  
v_p=S_w v_w+S_nv_n\nonumber\;,\\
v_m=S_w v'_w+S_nv'_n\;,
\end{eqnarray}
where $v'_w=dv_w/dS_w$ and $v'_n=dv_n/dS_w$.

But, what is the co-moving velocity $v_m$ physically? We first need to 
understand the thermodynamic velocities $\hat{v}_w$ and $\hat{v}_n$. These are
the velocities the two fluids would have had if they were miscible.  Equation
(\ref{eqn18}) then tells us that a change in the saturation $S_w$ leads to a 
change in the average seepage velocity $v_p$ which is the difference in
seepage velocities of the two fluids.  However, the two fluids are {\it not\/}
miscible and they do get in each other's way and the amount is dictated by the
co-moving velocity through equation (\ref{eqn18-1}). 

From equation (\ref{eqn18}) onwards to the end of this sections, none of
the equations contain the size of the REV.  If we now imagine a REV associated
with each point in the porous medium, we have a continuum description. We may 
then add equations that transport the fluids between these points.  Assuming that
the fluids are incompressible, these equations are       
\begin{equation}
\label{eqn5003}
\phi\frac{\partial S_w}{\partial t}=\frac{\partial\phi S_w v_w}{\partial x}\;,
\end{equation}
where $t$ is the time coordinate and $x$ is the spatial coordinate, and
\begin{equation}
\label{eqn5004}
\frac{\partial}{\partial x}\ \phi S_w =0\;.
\end{equation}
The generalization to three dimensions is straight forward.  

In order to connect the equations that now have been derived to a given
porous medium, constitutive equations for $v_p$ and $v_m$ need to be
supplied, linking the flow to the driving forces.  These may in the simplest
case be pressure gradient and saturation gradient.

%--------------------------------------------------------------------
\section{Differential transversal area distributions}
\label{secDiff}

In this section, we connect the pseudo-thermodynamic results of section
\ref{secThermo} to the properties of an underlying ensemble
distribution. This concept in the context of immiscible two-phase
flow was first considered by Savani et al.\ \cite{sbksvh17}. Here we
generalize this concept.  In some sense, we introduce here a statistical
mechanics from which the pseudo-thermodynamics ensue. 

We define a {\it differential transversal pore area\/} $a_p=a_p(S_w,v)$ 
where $v$ is a velocity such that $a_pdv$ is the pore area covered by fluid,
wetting or non-wetting, that has a velocity in the range $[v,v+dv]$.
Hence, $a_p$ --- and the other differential transversal pore areas that we
will proceed to construct --- are {\it statistical distributions of the pore
level velocities.\/}  The new idea we are introducing is that the velocity
distribution is measured in terms of transversal pore areas.  This makes it 
possible to make the connection between the flow at the pore level and the
pseudo-thermodynamic theory reviewed in the previous section. 

We must have that
\begin{equation}
\label{eqn3.1}
A_p=\int_{-\infty}^\infty dv\ a_p\;,
\end{equation}
where the integral runs over the entire range of negative and positive
velocities since there may be local areas where the flow direction is
opposite to the global flow. The total flow rate $Q$ is given by  
\begin{equation}
\label{eqn3.1}
Q=\int_{-\infty}^\infty dv\ v\ a_p\;,
\end{equation}
and the seepage velocity defined in equation
(\ref{eqn5.3}) is then given by
\begin{equation}
\label{eqn3.2}
v_p=\langle v\rangle_p=\frac{1}{A_p}\ \int_{-\infty}^\infty dv\ v\ a_p\;.
\end{equation}
Likewise, we define a wetting differential pore area $a_w$ and a
non-wetting differential pore area $a_n$. They have the same
properties except that they are restricted to the wetting or the
non-wetting fluids only.  That is,
\begin{equation}
\label{eqn3.10}
A_w=\int_{-\infty}^\infty dv\ a_w\;,
\end{equation}  
and
\begin{equation}
\label{eqn3.3}
A_n=\int_{-\infty}^\infty dv\ a_n\;.
\end{equation}
They relate to the wetting and non-wetting seepage velocities defined
in equations (\ref{eqn5.1}) and (\ref{eqn5.2}) as
\begin{equation}
\label{eqn3.4}
v_w=\langle v\rangle_w=\frac{1}{A_w}\ \int_{-\infty}^\infty dv\ v\ a_w\;,
\end{equation}
and
\begin{equation}
\label{eqn3.5}
v_n=\langle v\rangle_n=\frac{1}{A_n}\ \int_{-\infty}^\infty dv\ v\ a_n\;.
\end{equation}
We have that
\begin{equation}
\label{eqn3.6}
a_p=a_w+a_n\;.
\end{equation}
We now combine this equation with equation (\ref{eqn3.2}) to find
\begin{eqnarray}
\label{eqn3.7}
v_p&=&\frac{1}{A_p}\ \int_{-\infty}^\infty dv\ v\ \left(a_w+a_n\right)\nonumber\\
   &=&\left(\frac{A_w}{A_p}\right)\frac{1}{A_w}\int_{-\infty}^\infty dv\ 
   v\ a_w\nonumber\\
   &+&\left(\frac{A_n}{A_p}\right) 
   \frac{1}{A_n}\int_{-\infty}^\infty dv\ v\ a_n\nonumber\\
   &=&S_w v_w+S_n v_n\;,\\
   \nonumber
\end{eqnarray}
which is equation (\ref{eqn5.4}). We have here used equations
(\ref{eqn3}) and (\ref{eqn4}).

We may associate a differential area $a_m$ to the co-moving
velocity $v_m$ defined in equation (\ref{eqn18-1}). By using equations
(\ref{eqn3.2}), (\ref{eqn3.4}) and (\ref{eqn3.5}) in combination with
equation (\ref{eqn18-1}), we find
\begin{eqnarray}
\label{eqn3.8}
v_m&=&\frac{dv_p}{dS_w}-v_w+v_n\nonumber\\
   &=&\frac{1}{A_p}\int_{-\infty}^\infty dv\ v\ 
\left[\frac{\partial a_p}{\partial S_w}-\frac{a_w}{S_w}+\frac{a_n}{S_n}\right]\;,\nonumber\\
\end{eqnarray}
so that
\begin{eqnarray}
\label{eqn3.9}
a_m&=&\frac{\partial a_p}{\partial S_w}-\frac{a_w}{S_w}+\frac{a_n}{S_n}\nonumber\\
      &=&\left(\frac{\partial a_w}{\partial S_w}-\frac{a_w}{S_w}\right)
+\left(\frac{\partial a_n}{\partial S_w}+\frac{a_n}{S_n}\right)\;,\nonumber\\
\end{eqnarray}
where we have used equation (\ref{eqn3.6}). Equation (\ref{eqn3.9}) may be rewritten as
\begin{equation}
\label{eqn3.11}
a_m=S_w\frac{\partial }{\partial S_w}\left(\frac{a_w}{S_w}\right)
      +S_n\frac{\partial }{\partial S_w}\left(\frac{a_n}{S_n}\right)\;.
\end{equation}
Averaging this equation over $v$ and using equations (\ref{eqn3.4}),
(\ref{eqn3.5}) and (\ref{eqn3.8}) recovers equation (\ref{eqn20-1}). Hence, 
we note that equations (\ref{eqn3.9}) and (\ref{eqn3.11}) are the generalizations
of equations (\ref{eqn18-1}) and (\ref{eqn20-1}) to the differential transversal
areas. 

It follows that
\begin{equation}
\label{eqn3.100}
A_m=\int_{-\infty}^\infty dv\ a_m=0\;,
\end{equation}
where $A_m$ is the pore area associated with co-moving velocity $v_m$.
This is to expected as the areas $A_w$, $A_n$, $A_p$ and $A_m$ are ways to
partition the transversal pore area $A_p$; and we have that $A_w+A_n=A_p+0$. 
This implies that there is no volumetric flow rate associated with the co-moving
velocity since
\begin{equation}
\label{eqn3.12}
Q_m=A_mv_m=0\;.
\end{equation} 

Lastly, we may associate differential transversal areas to the thermodynamic
velocities defined in equations (\ref{eqn0.3000}) and (\ref{eqn0.3001}). We use
equations (\ref{eqn0.4}) and (\ref{eqn0.5}) to find
\begin{equation}
\label{eqn3.120}
\hat{a}_w=a_w+S_nS_w\ a_m\;,
\end{equation}
and 
\begin{equation}
\label{eqn3.121}
\hat{a}_n=a_n-S_wS_n\ a_m\;,
\end{equation}
where $a_m$ is given in equation (\ref{eqn3.11}). 
The thermodynamic velocities are then given by 
\begin{equation}
\label{eqn3.125}
\hat{v}_w=\frac{1}{A_w}\int_{-\infty}^\infty dv\ v\ \hat{a}_w\;,
\end{equation}
and
\begin{equation}
\label{eqn3.126}
\hat{v}_n=\frac{1}{A_n}\int_{-\infty}^\infty dv\ v\ \hat{a}_n\;.
\end{equation}

We find as expected that
\begin{equation}
\label{eqn3.122}
\hat{A}_w=\int_{-\infty}^\infty dv\ \hat{a}_w=A_w\;,
\end{equation}
and
\begin{equation}
\label{eqn3.123}
\hat{A}_n=\int_{-\infty}^\infty dv\ \hat{a}_n=A_n\;.
\end{equation}
Summing the two differential transversal areas for the thermodynamic
areas gives
\begin{equation}
\label{eqn3.124}
\hat{a}_w+\hat{a}_n=a_w+a_n=a_p\;.
\end{equation}
This leads us to an important remark. The differential transversal 
areas are statistical velocity distributions at the pore level.  
We see that the differential transversal areas that are associated 
with the thermodynamic velocities are different from those associated
with the seepage velocities.  However, equation (\ref{eqn3.124}) shows
that the {\it combined\/} differential transversal area based upon
the thermodynamic velocity distributions is the same as that based
upon the distributions giving the seepage velocities.  Hence, the 
two types of differential transversal areas represent a redistribution
of the pore level velocities, but in such a way that $A_w$ and $A_n$ are
preserved.  We see the same from equation (\ref{eqn3.100}) showing that
$A_m$ is zero and combining the equations (\ref{eqn3.120}) and (\ref{eqn3.121}).    

We see from equation (\ref{eqn3.9}) that $a_m$ is only zero if $a_w$
and $a_n$ are linear in $S_w$ and $S_n$ respectively, i.e., $a_w=S_w
b_w$ where $b_w$ is independent of $S_w$ and $a_n=S_n b_n$ where $b_n$
is independent of $S_n$. Hence, this is the condition for the
thermodynamic velocities to be equal to the seepage velocities.

%--------------------------------------------------------------------
\section{Moments and fluctuations}
\label{secMoments}

We define the $q$th moment of the seepage velocity distribution as
\begin{equation}
\label{eqn4.1}
v_p^q=\langle v^q\rangle_p=\frac{1}{A_p}\int_{-\infty}^{\infty}dv\ v^q a_p\;.
\end{equation}
By using equation (\ref{eqn3.6}) we find immediately
\begin{equation}
\label{eqn4.2}
v_p^q=v_w^q S_w+v_n^q S_n\;,
\end{equation}
where we have defined
\begin{equation}
\label{eqn4.3}
v_w^q=\langle v^q\rangle_w=\frac{1}{A_w}\ \int_{-\infty}^\infty dv\ v^q\ a_w\;,
\end{equation}
and
\begin{equation}
\label{eqn4.4}
v_n^q=\langle v^q\rangle_n=\frac{1}{A_n}\ \int_{-\infty}^\infty dv\ v^q\ a_n\;.
\end{equation}

We may work out the moments of the co-moving velocity are given by
\begin{equation}
\label{eqn4.22}
v_m^q=\frac{1}{A_p}\int_{-\infty}^{\infty}dv\ v^q\ a_m=
\left[\frac{dv_p^q}{dS_w}-v_w^q+v_n^q\right]\;,
\end{equation}
where we have used (\ref{eqn3.9}). 

The thermodynamic velocity moments may be defined as in a similar manner as the moments
of the seepage velocities, (\ref{eqn4.3}) and (\ref{eqn4.4}),
\begin{equation}
\label{eqn4.30}
\hat{v}_w^q=\langle \hat{v}^q\rangle_w=\frac{1}{A_w}\ \int_{-\infty}^\infty dv\ v^q\ \hat{a}_w\;,
\end{equation}
and
\begin{equation}
\label{eqn4.40}
\hat{v}_n^q=\langle \hat{v}^q\rangle_n=\frac{1}{A_n}\ \int_{-\infty}^\infty dv\ v^q\ \hat{a}_n\;.
\end{equation}
and we find
\begin{equation}
\label{eqn4.21}
\hat{v}_p^q=\hat{v}_w^q S_w+\hat{v}_n^q S_n\;,
\end{equation}
where we have used equations (\ref{eqn3.121}) and (\ref{eqn3.122}).

We may Fourier transform $a_p$, $a_w$ and $a_n$,
\begin{equation}
\label{eqn4.5}
2\pi\tilde{a}_p(\omega)=A_p\langle e^{i v\omega}\rangle_p
=\int_{-\infty}^\infty dv\ e^{iv\omega}\ a_p\;,
\end{equation}
\begin{equation}
\label{eqn4.6}
2\pi\tilde{a}_w(\omega)=A_w\langle e^{i v\omega}\rangle_w=\int_{-\infty}^\infty dv\ e^{iv\omega}\ a_w\;,
\end{equation}
and
\begin{equation}
\label{eqn4.7}
2\pi\tilde{a}_n(\omega)=A_n\langle e^{i v\omega}\rangle_n=\int_{-\infty}^\infty dv\ e^{iv\omega}\ a_n\;.
\end{equation}
From equation (\ref{eqn3.6}) we find
\begin{equation}
\label{eqn4.8}
\tilde{a}_p(S_w,\omega)=\tilde{a}_w(S_w,\omega)+\tilde{a}_n(S_w,\omega)\;,
\end{equation}
and
\begin{equation}
\label{eqn4.9} 
\langle e^{i v\omega}\rangle_p=S_w\langle e^{i v\omega}\rangle_w+S_n\langle e^{i v\omega}\rangle_n\;.
\end{equation}
We write $\langle \exp(iv\omega)\rangle_p$ as a cumulant expansion,
\begin{equation}
\label{eqn4.10}
\langle  e^{i v\omega}\rangle_p=\exp\left(\sum_{k=1}^\infty \frac{(i\omega)^k}{k!}\ C_p^k\right)\;,
\end{equation}
where $C_p^k$ is the $k$th cumulant. We define the wetting and
non-wetting velocity cumulants $C_w^k$ and $C_n^k$ in the same way. We
also write $\langle \exp(iv\omega)\rangle_p$ as a moment expansion
\begin{equation}
\label{eqn4.11}
\langle  e^{i v\omega}\rangle_p=\sum_{m=0}^\infty \frac{(i\omega)^m}{m!}\ v_p^m\;.
\end{equation}
By expanding the cumulant expression in equation (\ref{eqn4.10}) and
equating each power in $i\omega$ with the corresponding one in
equation (\ref{eqn4.11}), then repeating this for the wetting and
non-wetting cumulants, and lastly combining them through equation
(\ref{eqn4.9}), we find for the term proportional to $(i\omega)^2$,
\begin{equation}
\label{eqn4.12}
C_p^2+(C_p^1)^2=[C_w^2+(C_w^1)^2]S_w+[C_n^2+(C_n^1)^2]S_n\;.
\end{equation}
Noting that $C_p^1=v_p$, $C_w^1=v_w$ and $C_n^1=v_n$ and using the
notation $\Delta v_p^2=C_p^2$, $\Delta v_w^2=C_w^2$ and $\Delta
v_n^2=C_n^2$, we find from this equation
\begin{equation}
\label{eqn4.13}
\Delta v_p^2=\Delta v_w^2S_w+\Delta v_n^2S_n+S_wS_n\left(v_w-v_n\right)^2\;.
\end{equation}
We may follow this procedure for any of the cumulants.

The corresponding equation between the second cumulants of the
thermodynamic velocities is
\begin{equation}
\label{eqn4.14}
\Delta \hat{v}_p^2=\Delta \hat{v}_w^2S_w+\Delta \hat{v}_n^2S_n
+S_wS_n\left(\hat{v}_w-\hat{v}_n\right)^2\;.
\end{equation}
%--------------------------------------------------------------------
\section{Numerical observations}
\label{secNum}

The relations presented in Sections \ref{secDiff} and \ref{secMoments}
provide the bridge between the velocity distributions at the pore level
and the pseudo-thermodynamic theory outlined in Section \ref{secThermo}.
In order to test these relations, and to show how they may be used, we 
use a dynamic pore network simulator \cite{jh12}.

In pore network modeling, the porous medium is represented by a network
of pores which transport two immiscible fluids. The pore-network model
we consider here can be applied to regular networks such as a regular
lattice with an artificial disorder as well as to irregular networks
such as a reconstructed network from real samples. The flow of the two
immiscible fluids is described in this model by keeping the track of
all interface positions with time. This approach of pore network
modeling was first introduced by Aker et al.\ \cite{amhb98}
for drainage displacements in a regular network. Over the last two
decades, new mechanisms have been developed to extend the model for
the steady-state flow as well as for irregular networks. A detailed
description of this model in its most recent form can be found in
\cite{gvkh19,gwh19,sgvh19} and we therefore describe it here only briefly.

%--------------------------------------------------------------------
\begin{figure*}[ht]
  \includegraphics[width=0.3\textwidth,clip]{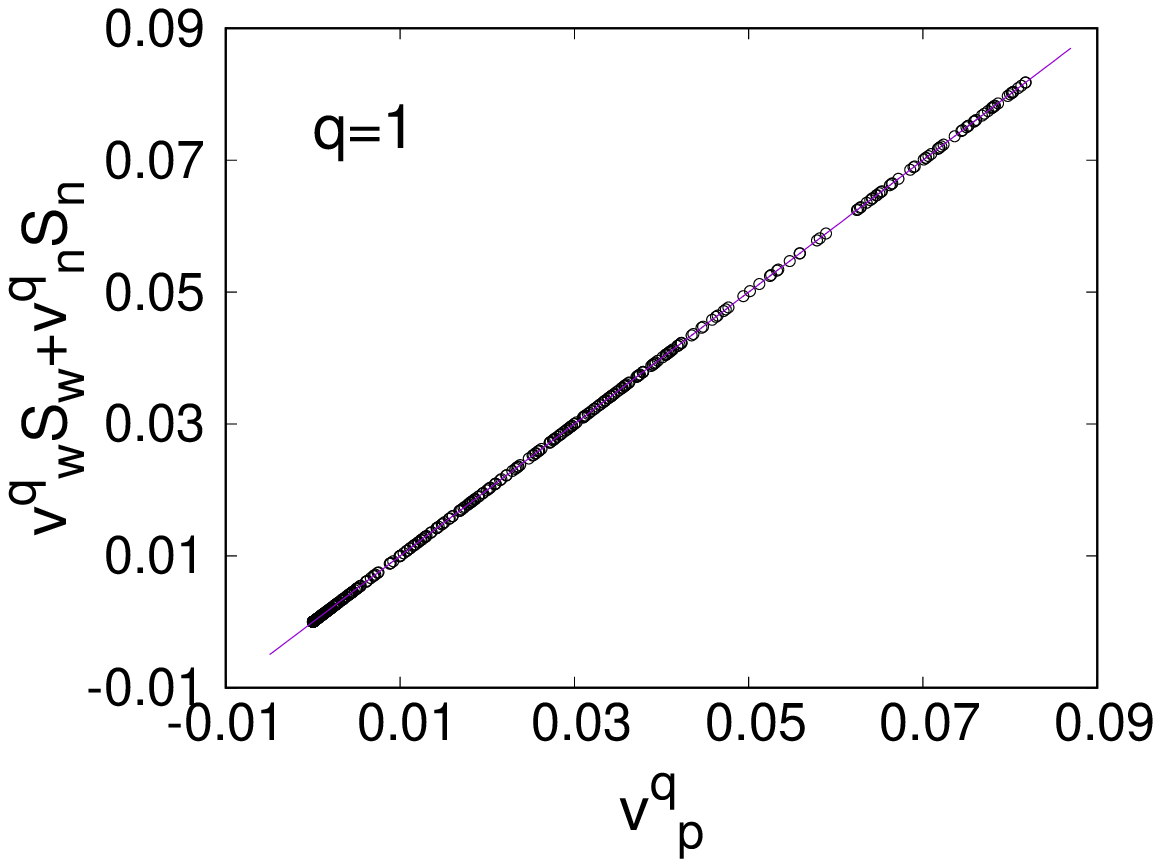}
  \includegraphics[width=0.3\textwidth,clip]{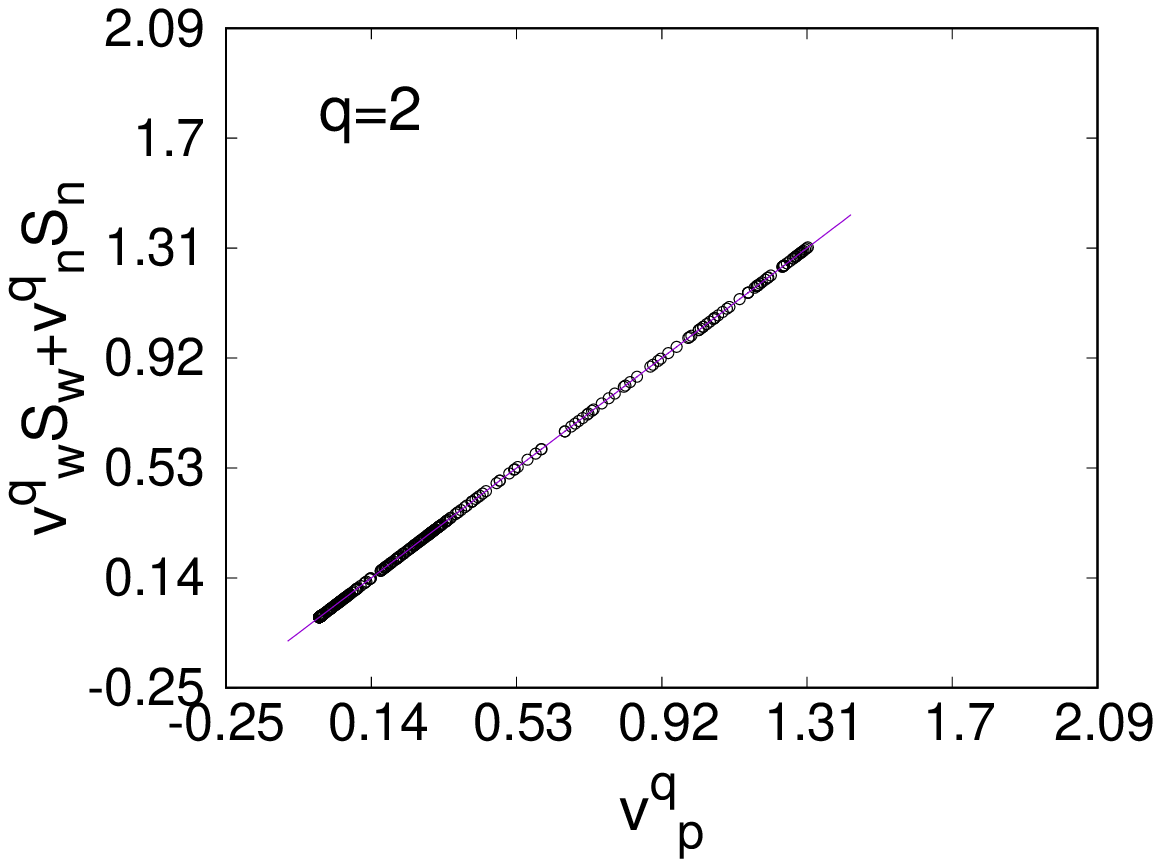}
  \includegraphics[width=0.3\textwidth,clip]{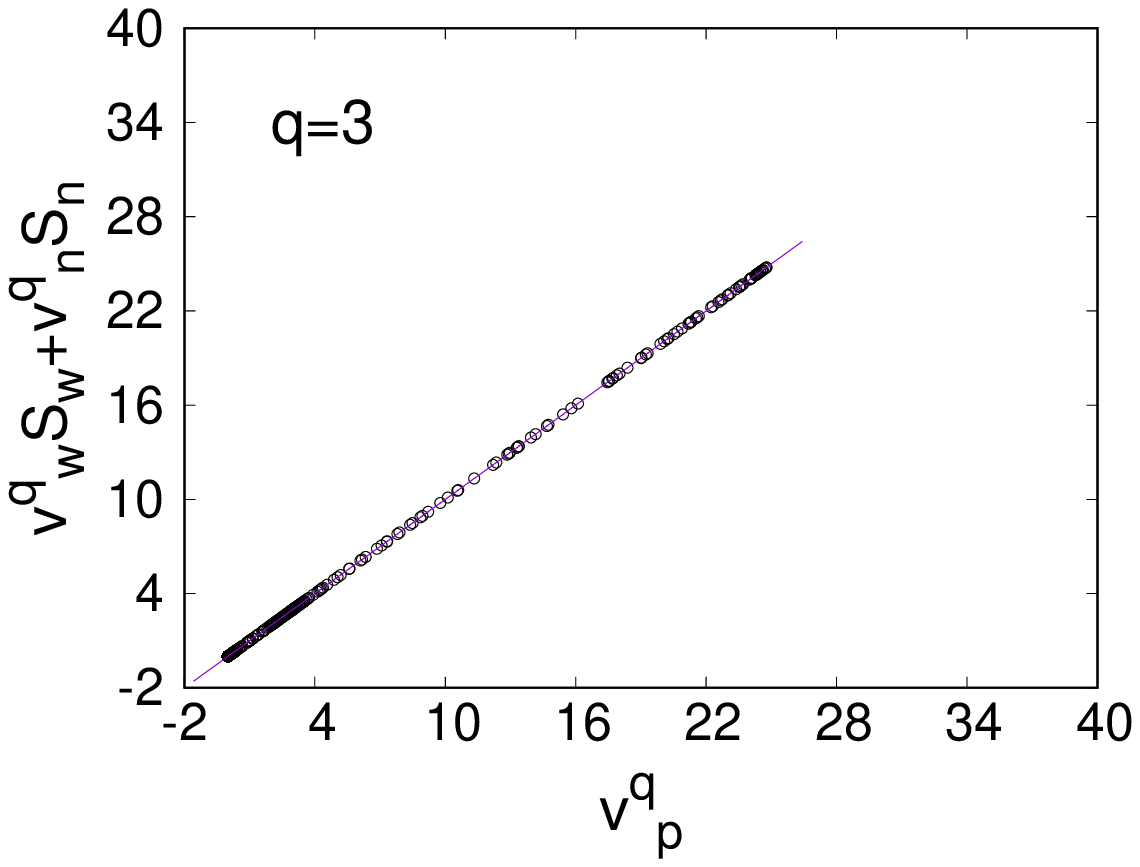}
  \includegraphics[width=0.3\textwidth,clip]{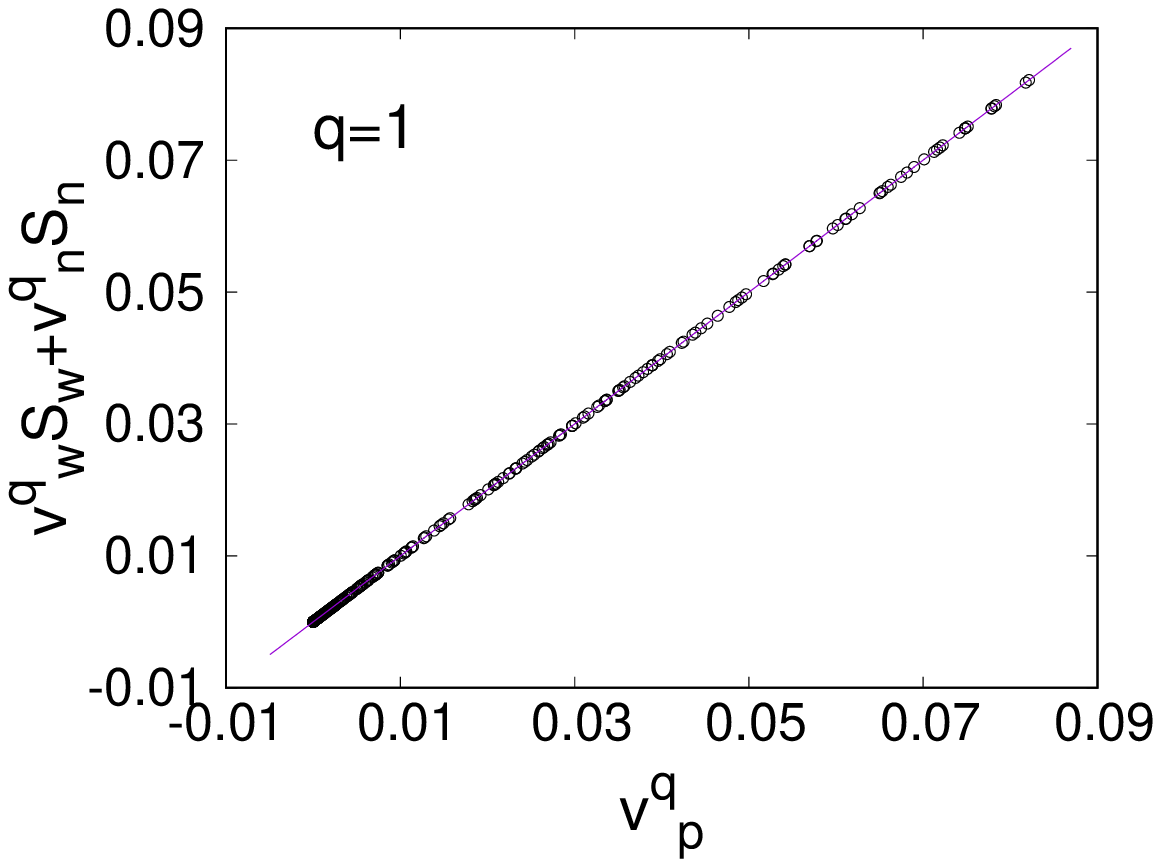}
  \includegraphics[width=0.3\textwidth,clip]{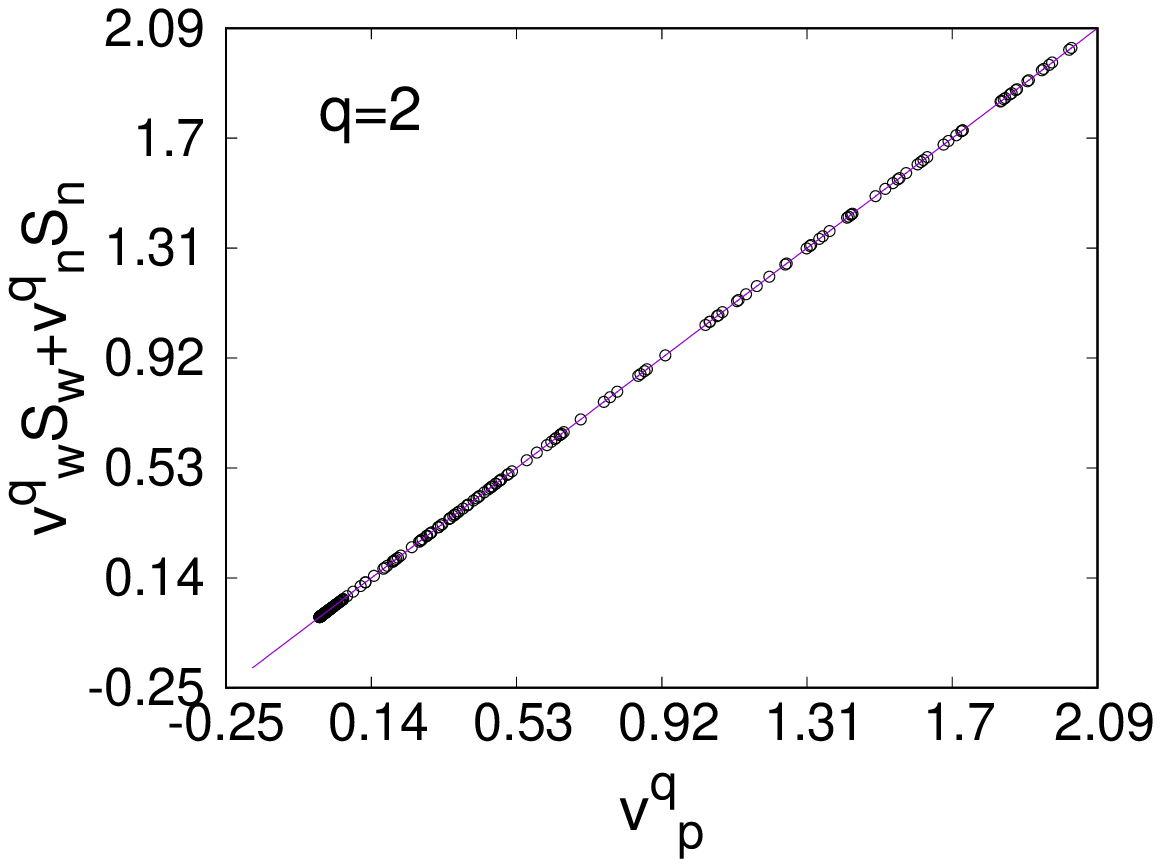}
  \includegraphics[width=0.3\textwidth,clip]{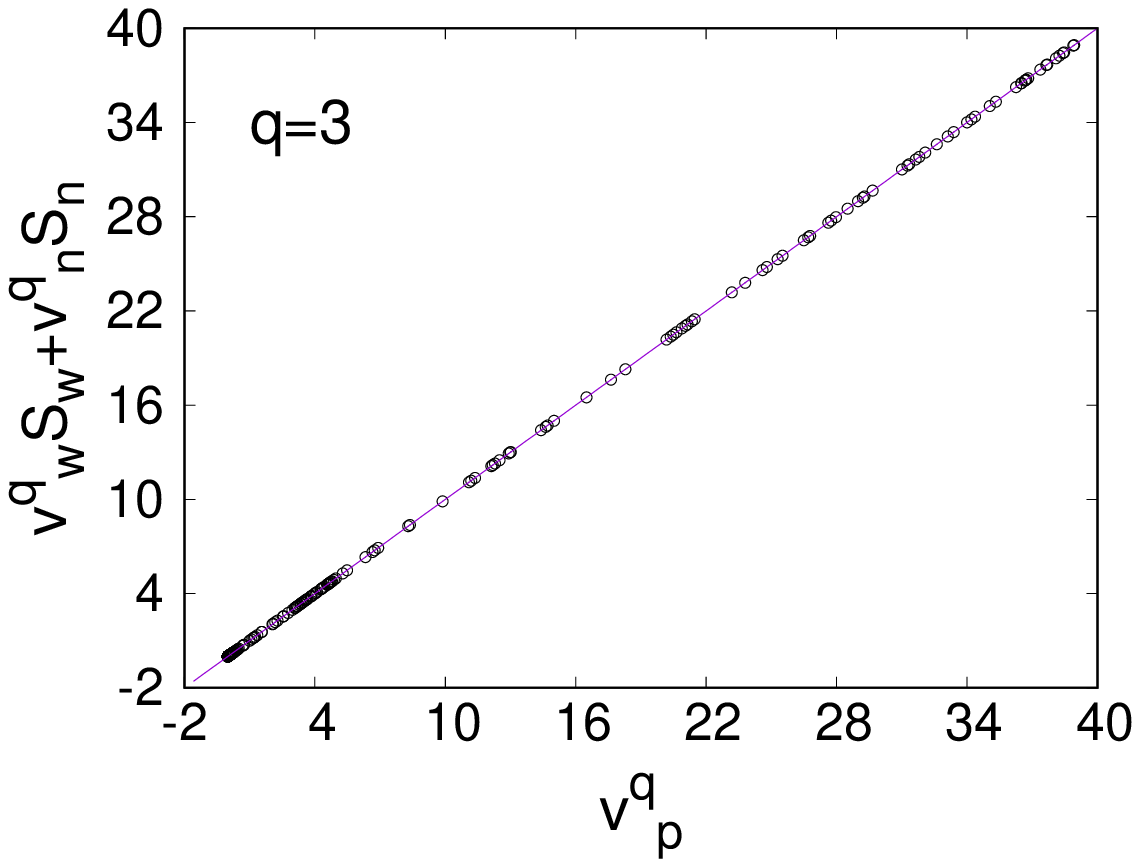} 
  \caption{Verification of the relations (\ref{eqn5.4}),
    (\ref{eqn3.7}) and (\ref{eqn4.2}) between the steady-state seepage
    velocities $v_p$, $v_w$ and
    $v_n$, and their higher moments for the 2D
    regular network. The top row represents the direct approach of
    measurements using equations (\ref{eqn5.03}), (\ref{eqn5.04}) and
    (\ref{eqn5.05}). The bottom row corresponds to the velocities
    measured from the differential area distributions defined in
    equations (\ref{eqn4.1}), (\ref{eqn4.3}) and (\ref{eqn4.4}).}
\label{fig1}
\end{figure*}
%--------------------------------------------------------------------

%--------------------------------------------------------------------
\begin{figure*}[ht]
  \includegraphics[width=0.31\textwidth,clip]{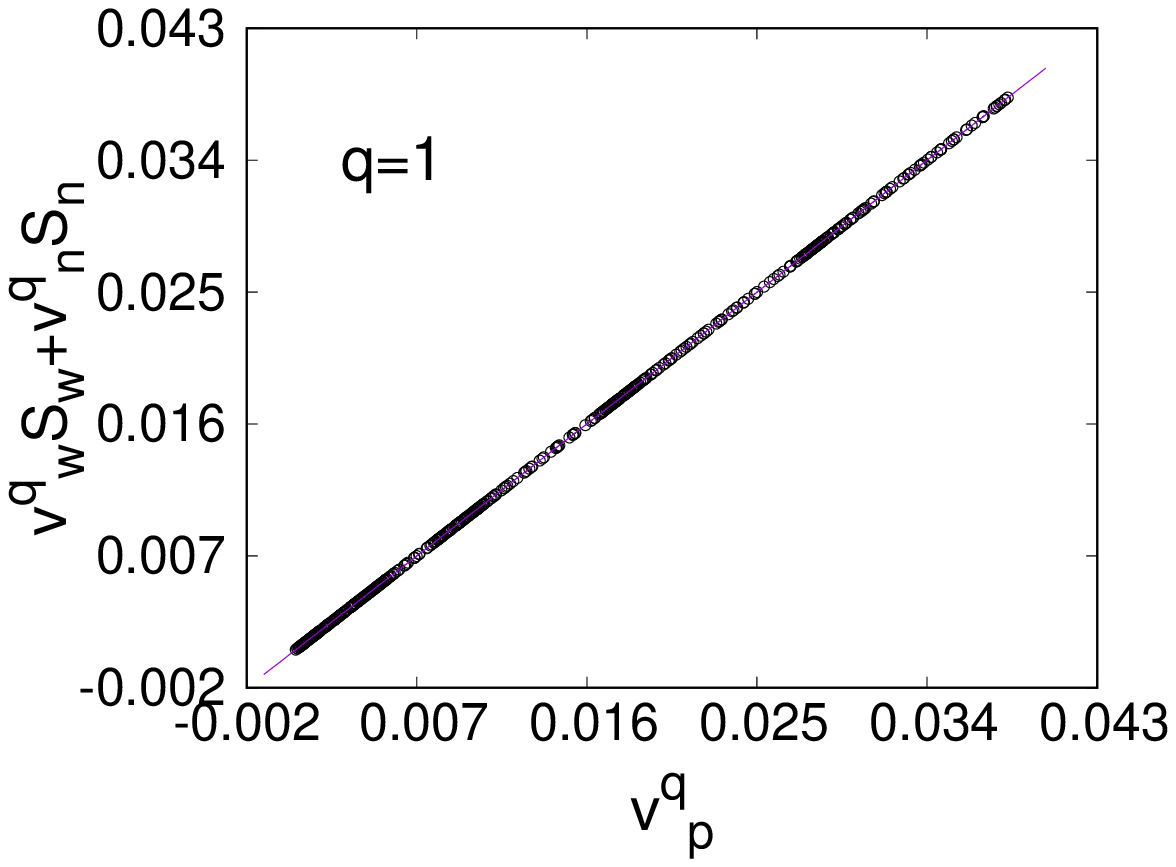}
  \includegraphics[width=0.3\textwidth,clip]{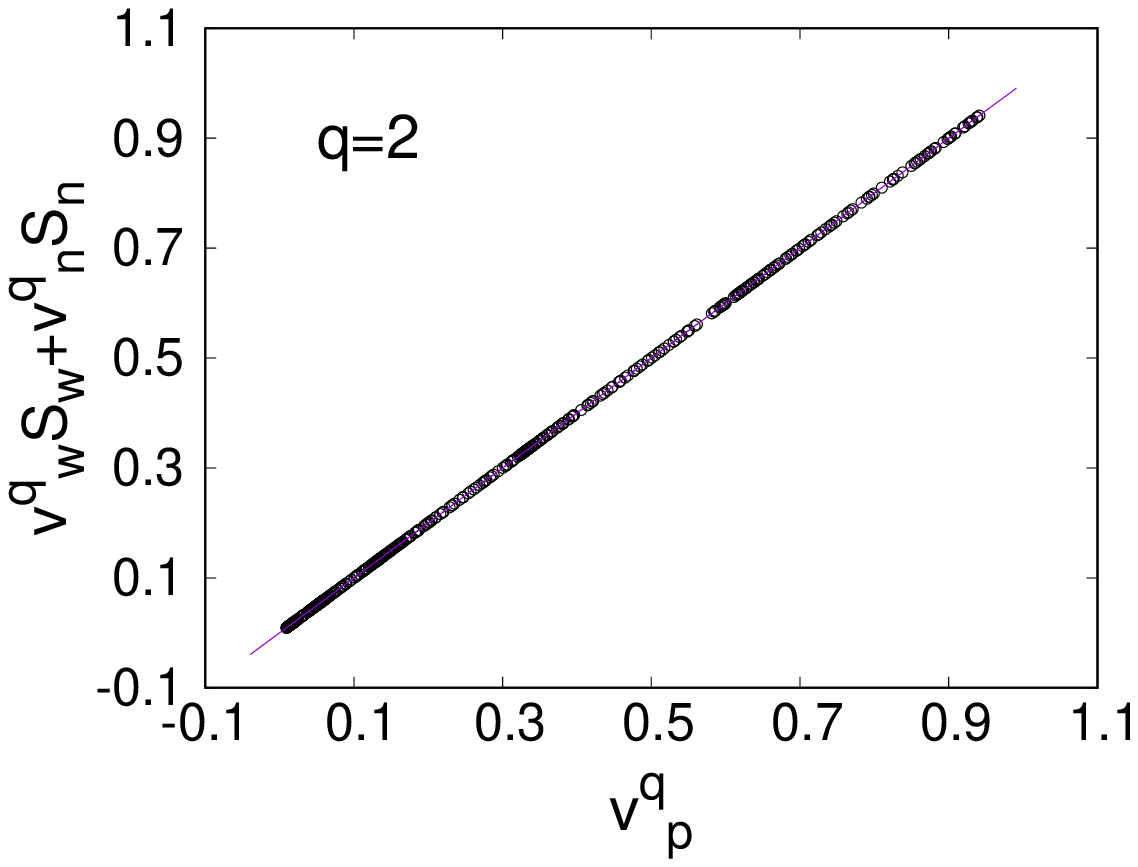}
  \includegraphics[width=0.29\textwidth,clip]{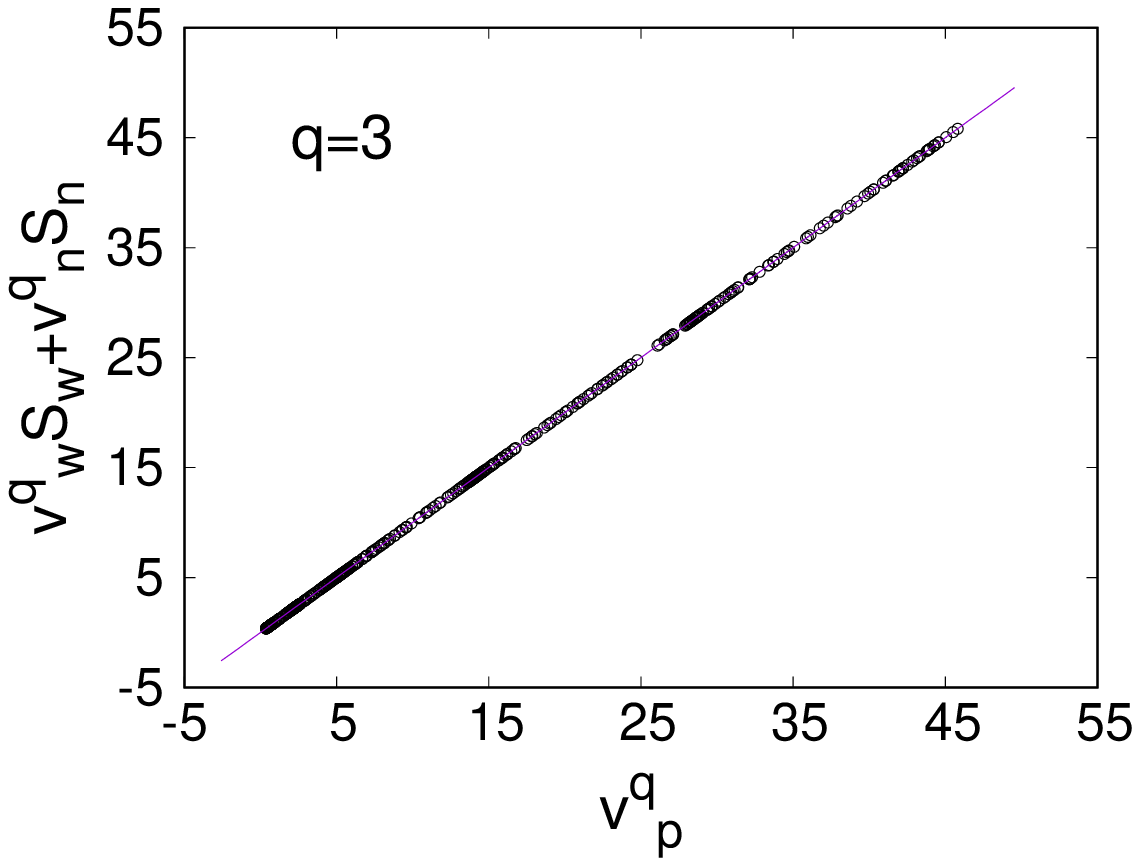}
  \includegraphics[width=0.31\textwidth,clip]{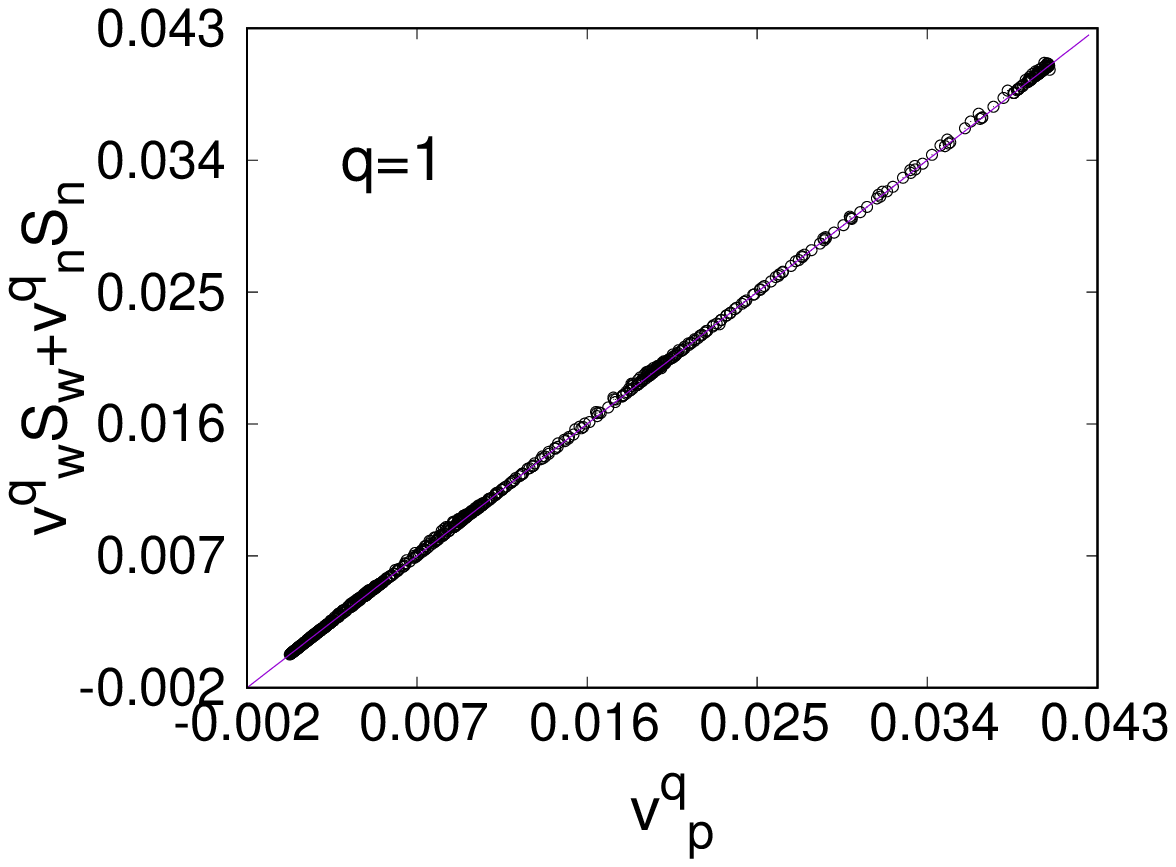}
  \includegraphics[width=0.3\textwidth,clip]{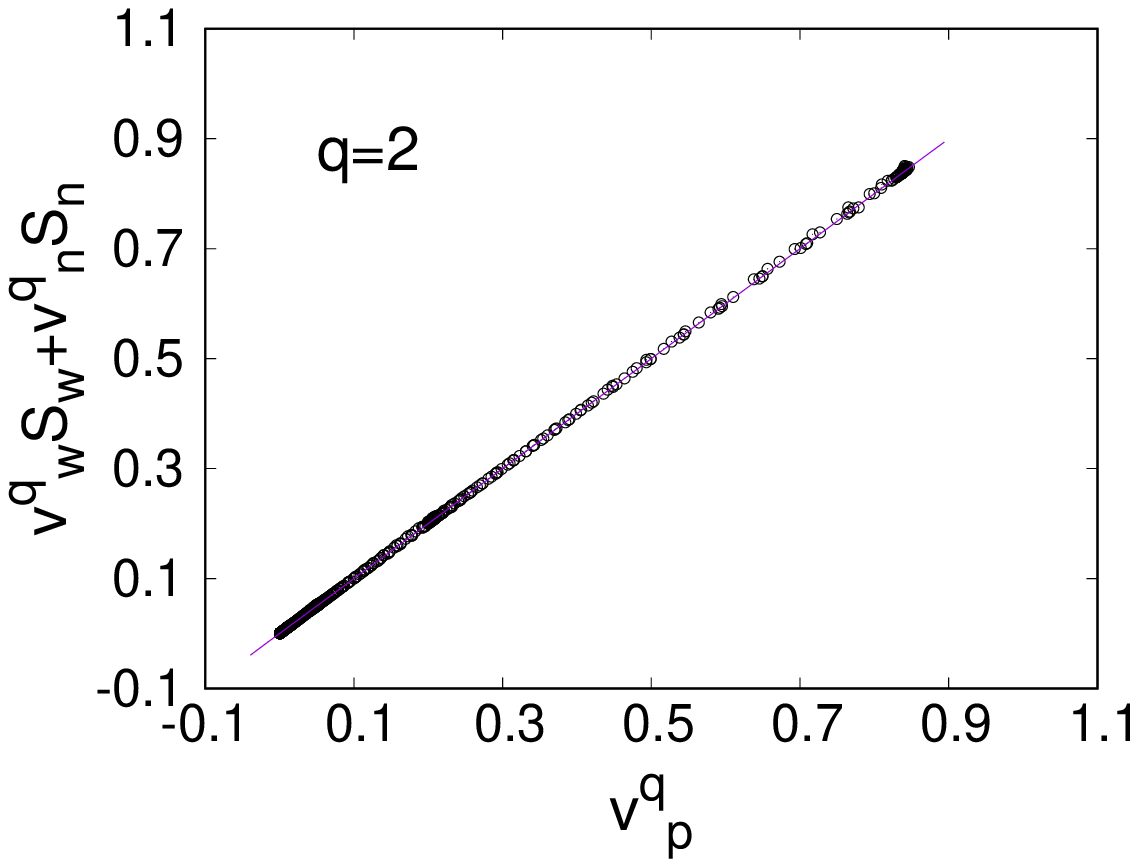}
  \includegraphics[width=0.3\textwidth,clip]{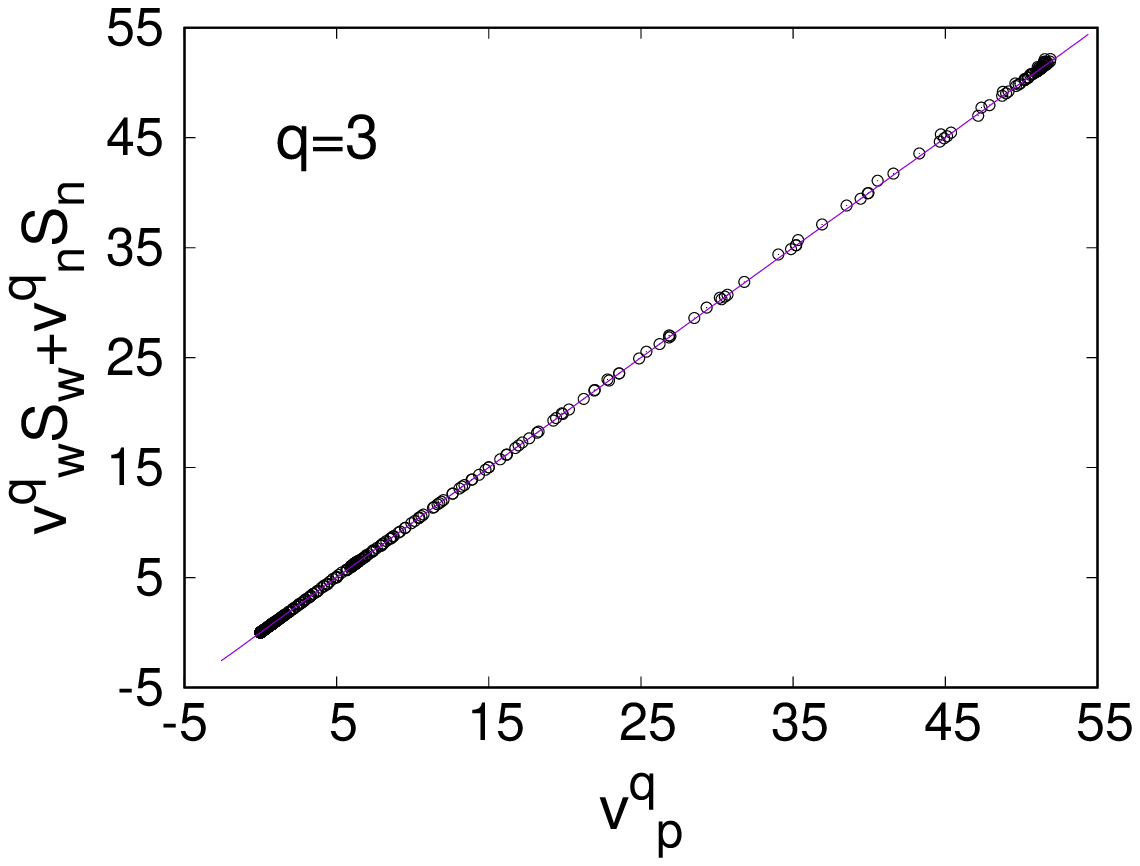} 
  \caption{Verification of the relations (\ref{eqn5.4}),
    (\ref{eqn3.7}) and (\ref{eqn4.2}) between the steady-state seepage
    velocities $v_p$, $v_w$ and
    $v_n$, and their higher moments for the 3D Berea
    network. The direct approach of measurements using equations
    (\ref{eqn5.03a})--(\ref{eqn5.05a}) are presented in the top
    row. The measurements using the differential area distributions
    defined in equations (\ref{eqn4.1}), (\ref{eqn4.3}) and
    (\ref{eqn4.4}) are presented in the bottom row.}
\label{fig2}
\end{figure*}
%--------------------------------------------------------------------

The porous medium is represented by a network of links that are
connected at nodes. All the pore space in this model is assigned to
the links and, hence, the nodes do not contain any volume, they only represent
the positions where the links meet. The flow rate $q_j$ inside any
link $j$ of the network at any instant of time for fully developed
viscous flow is obtained by \cite{w21,shbk13},
\begin{equation}
  \displaystyle
  q_j = -\frac{g_j}{l_j\mu_j}\left[\Delta p_j - \sum p_{c,j}\right]
  \label{eqnWB}
\end{equation}
where $\Delta p_j$ is the pressure drop across link, $l_j$ is the
link length and $g_j$ is the link mobility which depends on the cross
section of the link. The viscosity term $\mu_j$ is the
saturation-weighted viscosity of the fluids inside the link given by
$\mu_j = s_{j,w}\mu_w + s_{j,n}\mu_n$ where $\mu_w$ and $\mu_n$ are
the wetting and non-wetting viscosities and $s_{j,w}$ and $s_{j,n}$
are the wetting and non-wetting fluid saturations inside the link
respectively. The term $\sum p_{c,j}$ corresponds to the sum of all
the interfacial pressures inside the $j$th link. A pore typically
consists of two wider pore bodies connected by a narrow pore
throat. We model this by using hour-glass shaped links. The variation
of the interfacial pressure with the interface position for such a
link is modeled by \cite{shbk13},
\begin{equation}
  \displaystyle
  |p_c\left(x\right)| = \frac{2\gamma\cos\theta}{r_j}
  \left[1-\cos\left(\frac{2\pi x}{l_j}\right)\right]
  \label{eqnpc}
\end{equation}
where $r_j$ is the average radius of the link and $x \in [0,l_j]$ is
the position of the interface inside the link. Here $\gamma$ is the
surface tension between the fluids and $\theta$ is the contact angle
between the interface and the pore wall. These two equations
(\ref{eqnpc}) and (\ref{eqnWB}), together with the Kirchhoff
relations, that is, the sum of the net volume flux at every node at
each time step will be zero, provide a set of linear equations. We
solve these equations with conjugate gradient solver \cite{bh88} to
calculate the local flow rates. All the interfaces are then advanced
accordingly with small time steps. In order to achieve steady-state
flow, we apply periodic boundary conditions in the direction of flow.

We construct a diamond lattice with $64 \times 64$ links in two
dimensions (2D) with link lengths $l_j = 1\,{\rm mm}$ for each
link. Disorder is introduced by choosing the link radii $r_j$ randomly
from a uniform distribution in the range $0.1\,{\rm mm}$ and
$0.4\,{\rm mm}$. We use $10$ different realizations of such network
for our simulations in 2D. In three dimensions (3D), we use a network
reconstructed from a $1.8\times 1.8\times 1.8\,{\rm mm}^3$ sample of
Berea sandstone that contains $2274$ links and $1163$ nodes
\cite{rho09}. Simulations are performed under constant pressure drop
$\Delta P$ across the network. For 2D, we have considered $3$
different values for pressure pressure drop such that, $\Delta
P/L=0.5$, $1.0$ and $1.5\,{\rm MPa/m}$. For the 3D network, values
of $\Delta P/L$ are chosen as, $10$, $20$, $40$ and $80\,{\rm
MPa/m}$. The values for surface tension $\gamma$ are chosen to
be $0.02$, $0.03$ and $0.04\,{\rm N/m}$ for both 2D and 3D. Three
different values of viscosity ratios $M(=\mu_n/\mu_w) = 0.5$, $1.0$
and $2.0$ are considered. These values are chosen in such a way that
the capillary number, defined as
\begin{equation}
\label{eqn5.01}
{\rm Ca} = \displaystyle\frac{\mu_eQ}{\gamma A_p}
\end{equation}
falls in a range of around $10^{-3}$ to $10^{-1}$. Here $\mu_e$ is the
saturation weighted effective viscosity of the system given by $\mu_e
= S_w\mu_w + S_n\mu_n$. Specifically, we find ${\rm Ca}$ in the range
of $0.004$ to $0.074$ for 2D and $0.001$ to $0.271$ for 3D in the steady
state. As the simulations are performed under constant pressure drop,
the capillary number fluctuates. $\rm Ca$ is therefore calculated as
functions of time by measuring the total flow rate $Q$ along any cross
section of the network perpendicular to the applied pressure drop. For
any set of parameters, saturations are varied in the steps of $0.05$
from $0$ to $1$ which correspond to $21$ saturation values.

The simulations are continued to the steady state which is
defined by the global measurable quantities, such as the
fractional flow or the total flow rate $Q$ fluctuate around a steady
average. In the steady state, we calculate the seepage velocities
averaged over time. First we use direct measurements, where we
measure the global flow rates ($Q$, $Q_w$ and $Q_n$) and the pore
areas ($A_p$, $A_w$ and $A_n$) through any cross section orthogonal to
the applied pressure drop and then use equations
(\ref{eqn5.1})--(\ref{eqn5.3}) to calculate the seepage
velocities. Next, we perform the measurements using the differential
pore areas ($a_p$, $a_w$ and $a_n$) and calculate the seepage
velocities using description given in section \ref{secDiff}. We then
compare the results from the two measurements and calculate the
co-moving velocities. We then verify the relation between the seepage
velocities and their higher moments.
%--------------------------------------------------------------------
\begin{figure}
  \centerline{\hfill\includegraphics[width=0.4\textwidth,clip]{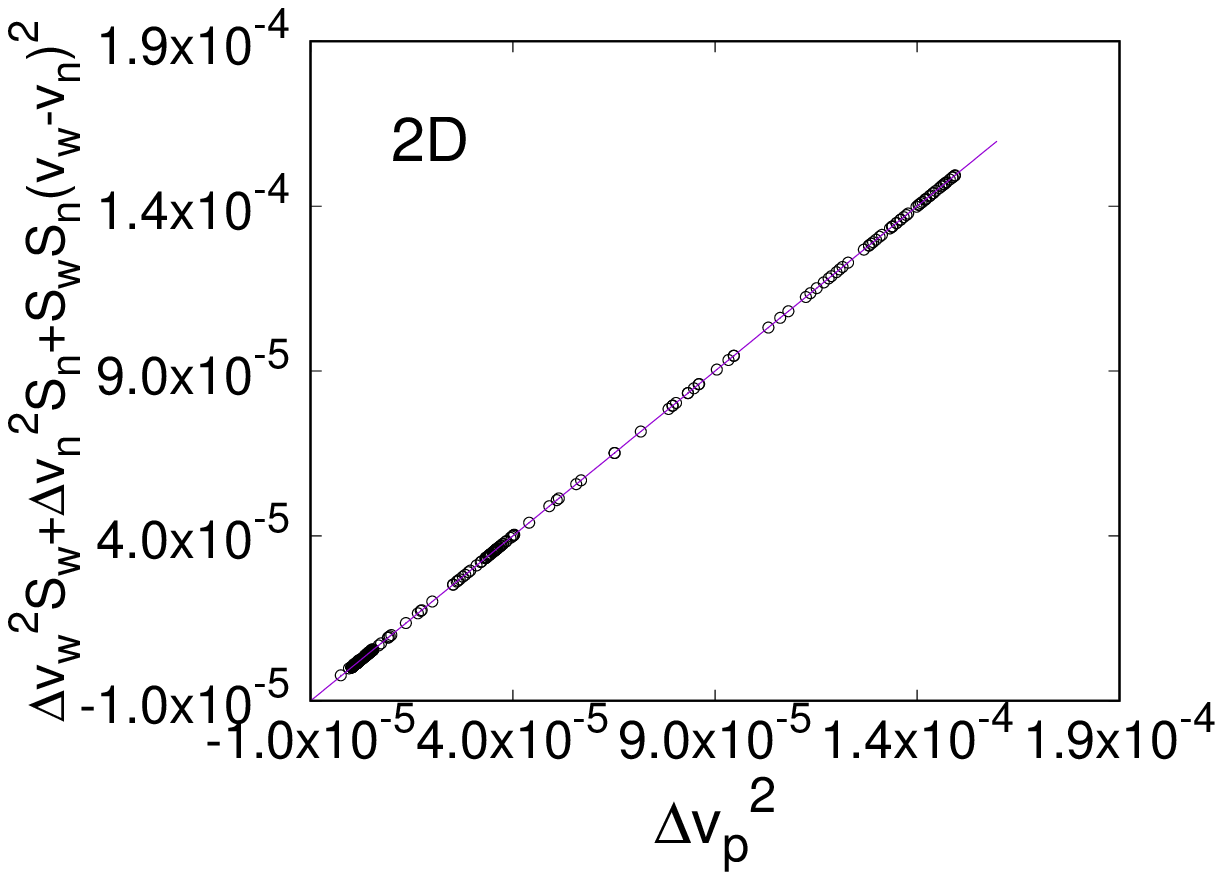}\hfill}
  \centerline{\hfill\includegraphics[width=0.4\textwidth,clip]{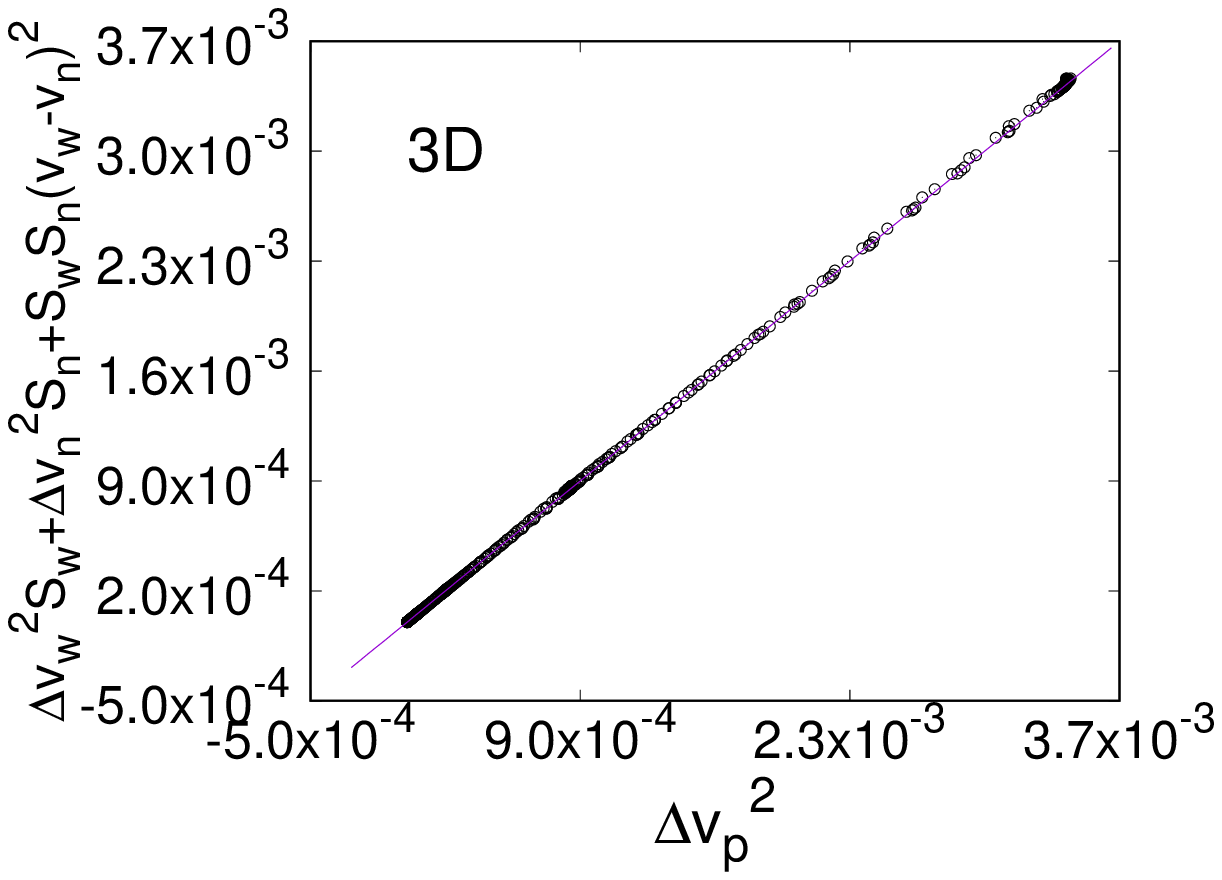}\hfill}
  \caption{Numerical verification of equation (\ref{eqn4.13}) between
    the fluctuations in the seepage velocities.}
\label{fig3}
\end{figure}
%--------------------------------------------------------------------

For the direct measurements, imagine a cross section at any place of
the network orthogonal to the overall direction of flow. For
the regular diamond lattice in 2D, all the links have the same
length. Different moments of the seepage velocities can therefore be
calculated by
\begin{equation}
  \label{eqn5.03}
  \displaystyle
  v^q_p = \frac{\displaystyle\sum_j\left(\frac{q_j}{a_j}\right)^q a_j}{\displaystyle\sum_ja_j}\;,
\end{equation}
\begin{equation}
  \label{eqn5.04}
  \displaystyle
  v^q_w = \frac{\displaystyle\sum_j\left(\frac{q_j}{a_j}\right)^q 
  a_jS_{w,j}}{\displaystyle\sum_ja_jS_{w,j}}\;,
\end{equation}
and
\begin{equation}
  \label{eqn5.05}
  \displaystyle
  v^q_n = \frac{\displaystyle\sum_j\left(\frac{q_j}{a_j}\right)^q 
  a_jS_{n,j}}{\displaystyle\sum_ja_jS_{n,j}}\;,
\end{equation}
where $a_j$ is the projection of the pore area of the $j$th link on
the the cross sectional plane. Here, all links have the same angle
$\alpha=45^{\circ}$ with the direction of the overall flow. However,
in case of the irregular network in 3D, equations
(\ref{eqn5.03})--(\ref{eqn5.05}) need to be modified as the links have
different lengths and orientations. In such case, the one can
calculate the seepage velocities by \cite{sgvh19},
\begin{equation}
  \label{eqn5.03a}
  \displaystyle
  v^q_p = \frac{\displaystyle\sum_j\left(\frac{q_j}{a_j}\right)^q 
  a_jl_{x,j}}{\displaystyle\sum_ja_jl_{x,j}}\;,
\end{equation}
\begin{equation}
  \label{eqn5.04a}
  \displaystyle
  v^q_w = \frac{\displaystyle\sum_j\left(\frac{q_j}{a_j}\right)^q 
  a_jS_{w,j}l_{x,j}}{\displaystyle\sum_ja_jS_{w,j}l_{x,j}}\;,
\end{equation}
and
\begin{equation}
  \label{eqn5.05a}
  \displaystyle
  v^q_n = \frac{\displaystyle\sum_j\left(\frac{q_j}{a_j}\right)^q 
  a_jS_{n,j}l_{x,j}}{\displaystyle\sum_{j}a_jS_{n,j}l_{x,j}}\;,
\end{equation}
where $l_{x,j}=l_j\cos\alpha_j$ is the projection of the link length ($l_j$) to the direction of the overall flow. 

If we in these equations consider every link having the same length $l_j=l$ and same orientations $\alpha_j=\alpha$, we retrieve the equations (\ref{eqn5.03})--(\ref{eqn5.05}). 
For the first moment ($q=1$), the velocities $v^q_p$, $v^q_w$ and $v^q_n$ are equivalent to $Q/A_p$, 
$Q/A_w$ and $Q/A_n$ respectively in both 2D and 3D. 
%--------------------------------------------------------------------
\begin{figure*}[ht]
  \includegraphics[width=0.29\textwidth,clip]{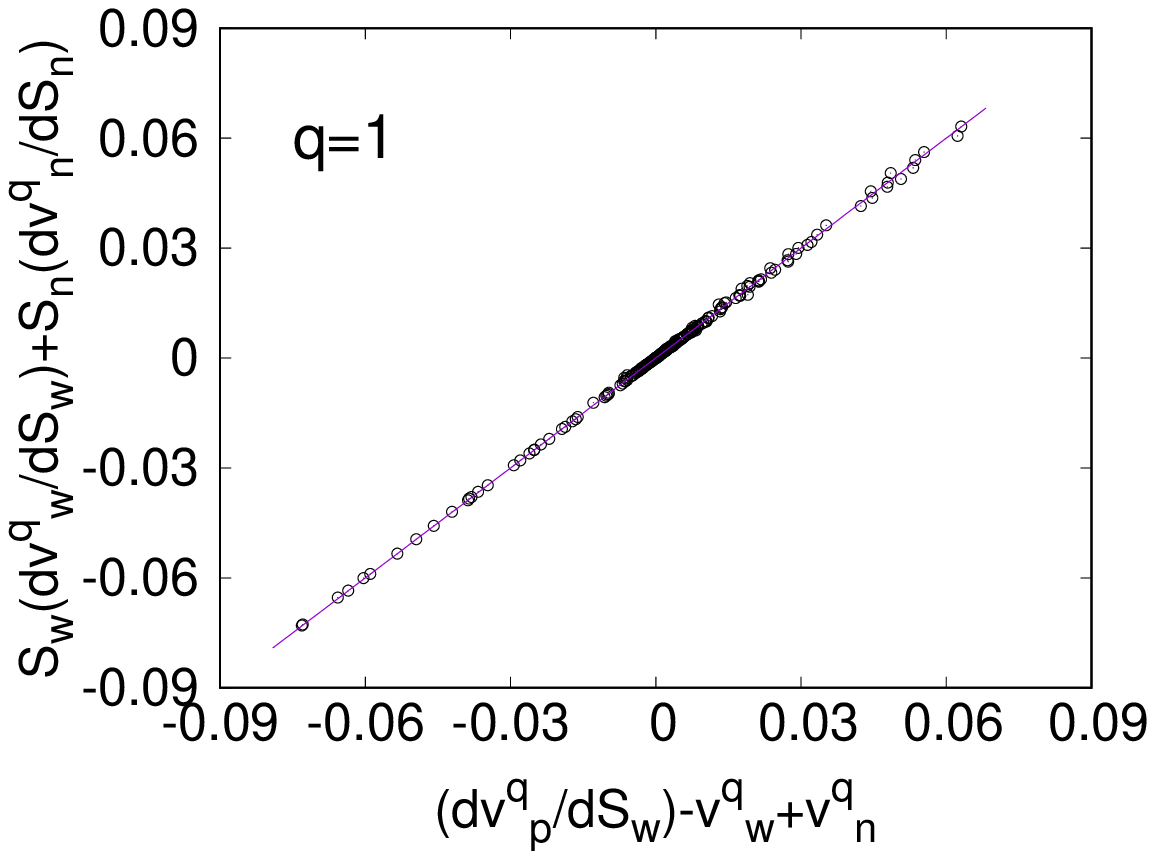}
  \includegraphics[width=0.3\textwidth,clip]{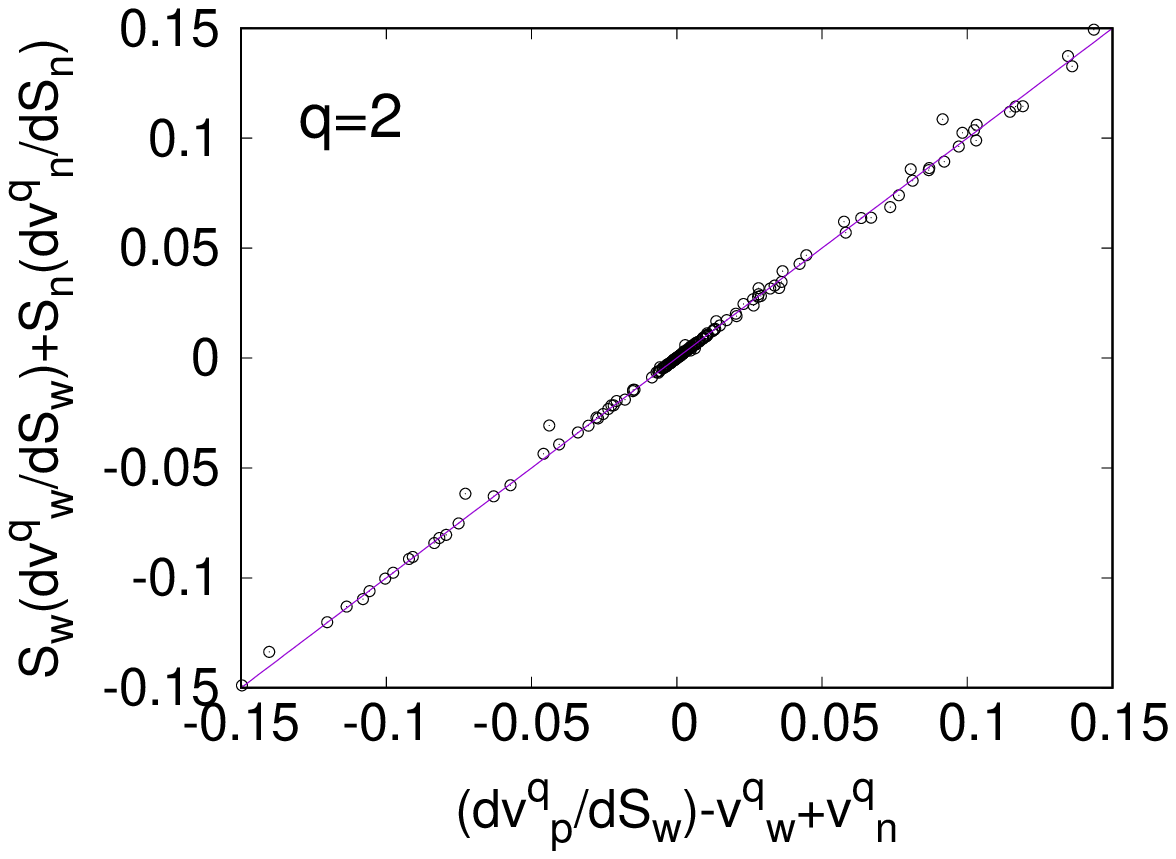}
  \includegraphics[width=0.29\textwidth,clip]{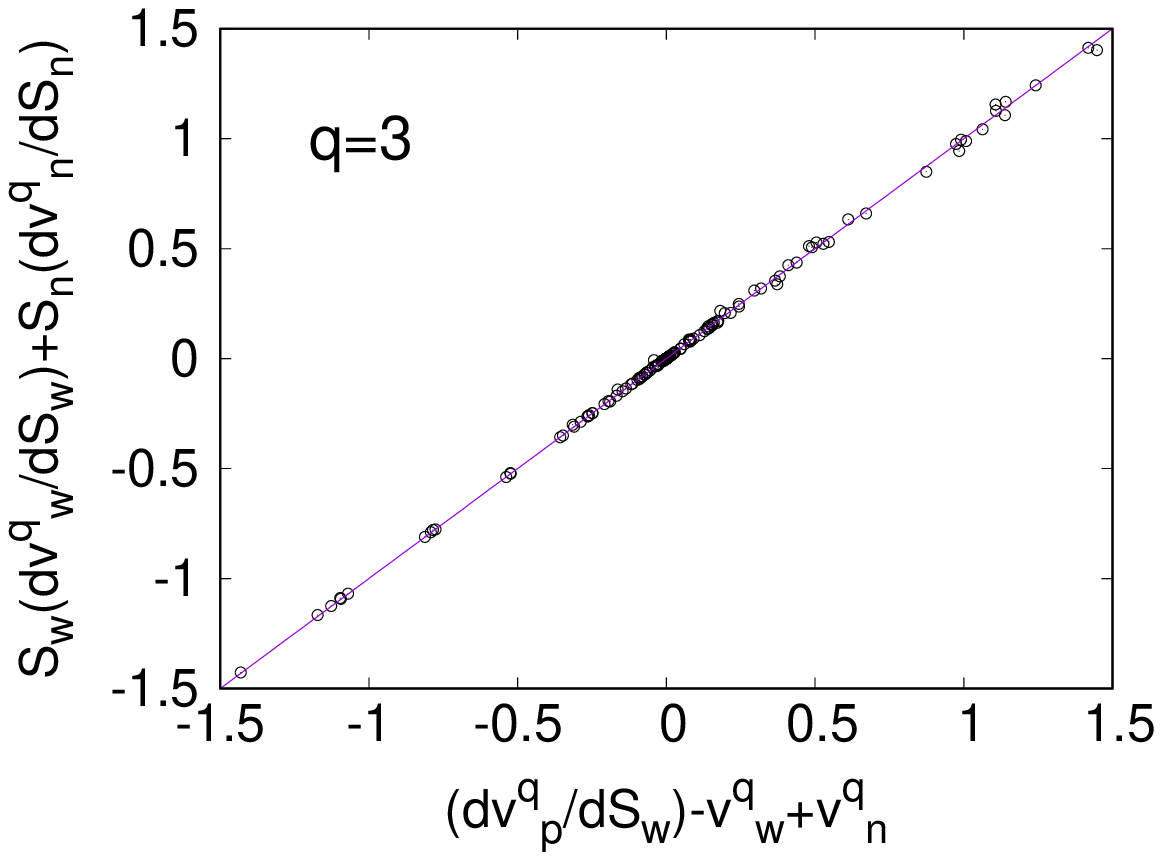}
  \includegraphics[width=0.3\textwidth,clip]{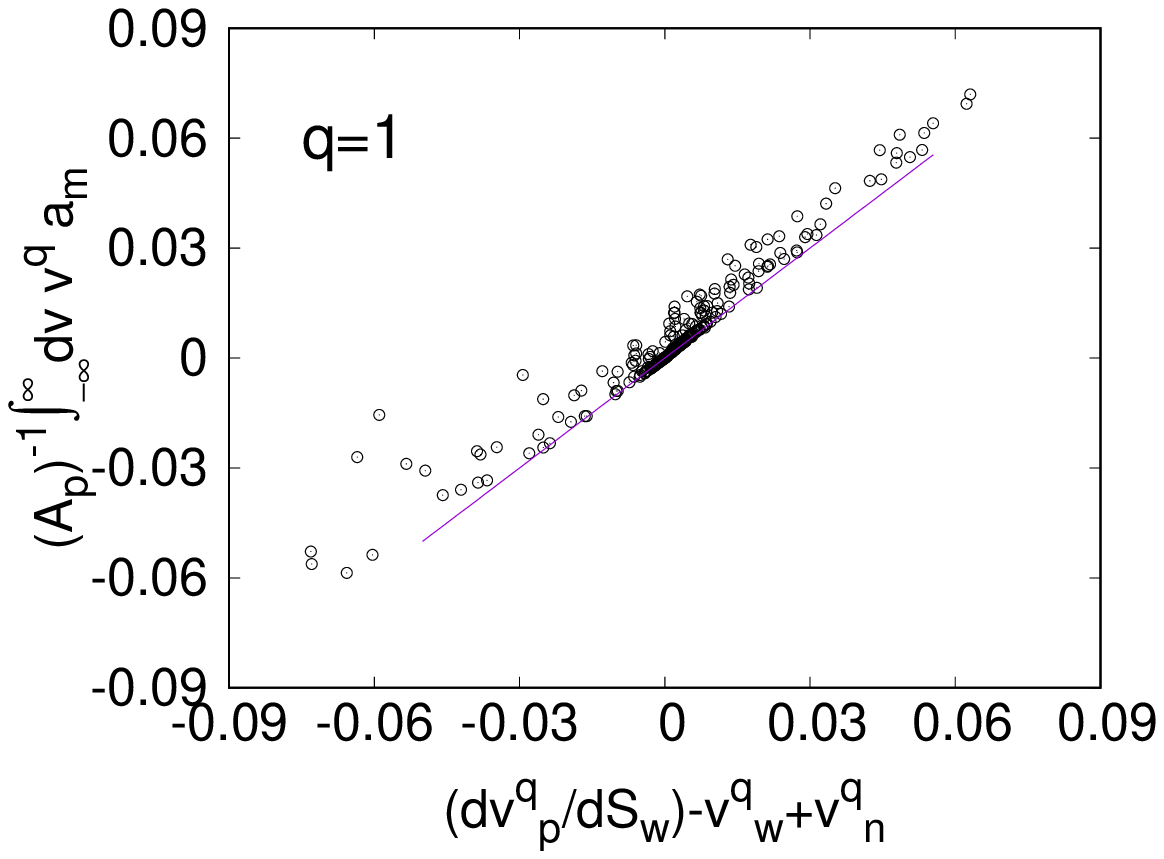}
  \includegraphics[width=0.3\textwidth,clip]{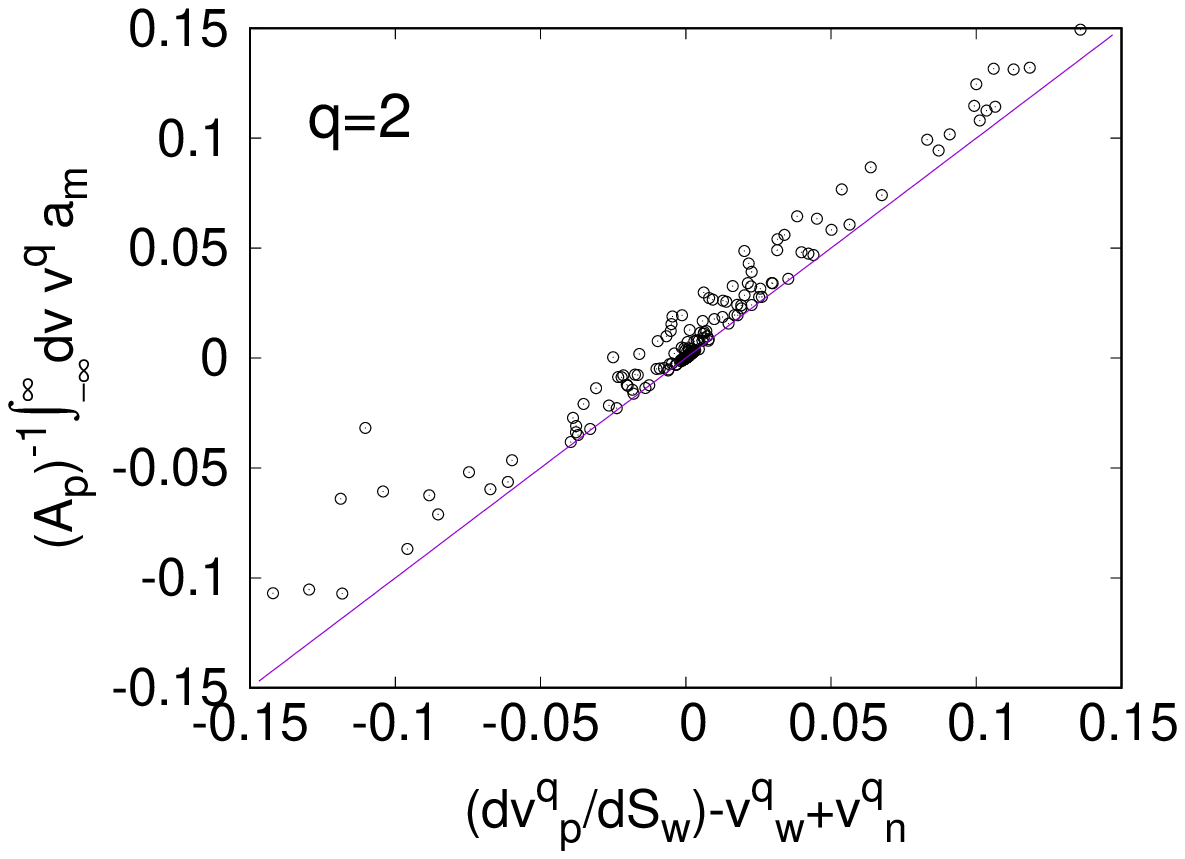}
  \includegraphics[width=0.29\textwidth,clip]{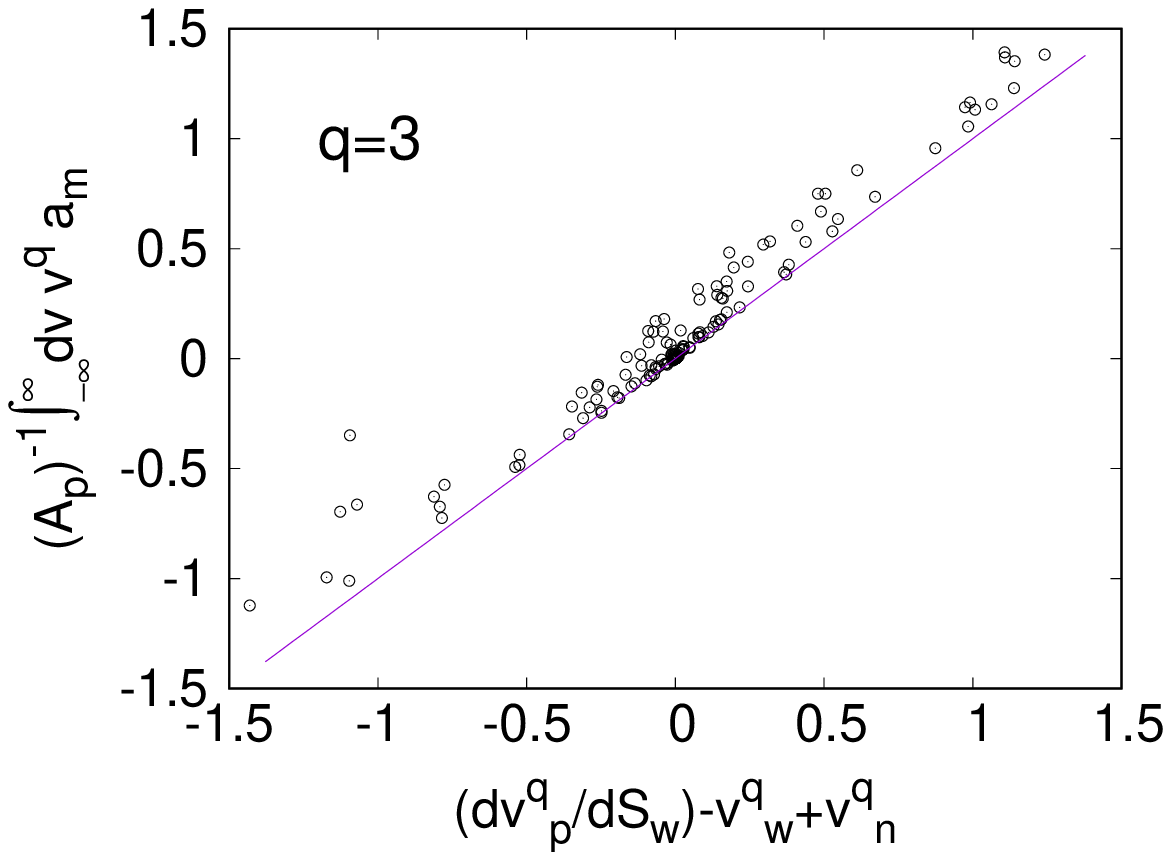} 
\caption{Measurement of the co-moving velocity ($v_m$) and its heigher
  moments for the 2D network. The top row corresponds to the
  calculations using equations (\ref{eqn18-1}) and (\ref{eqn20-1})
  with the direct measurements. The bottom row shows the measurements
  of $v_m^q$ using the differential area distributions with equation
  (\ref{eqn3.8}) and compared with the direct measurements where higher
  fluctuations are observed.}
\label{fig4}
\end{figure*}
\begin{figure*}[ht]
  \includegraphics[width=0.29\textwidth,clip]{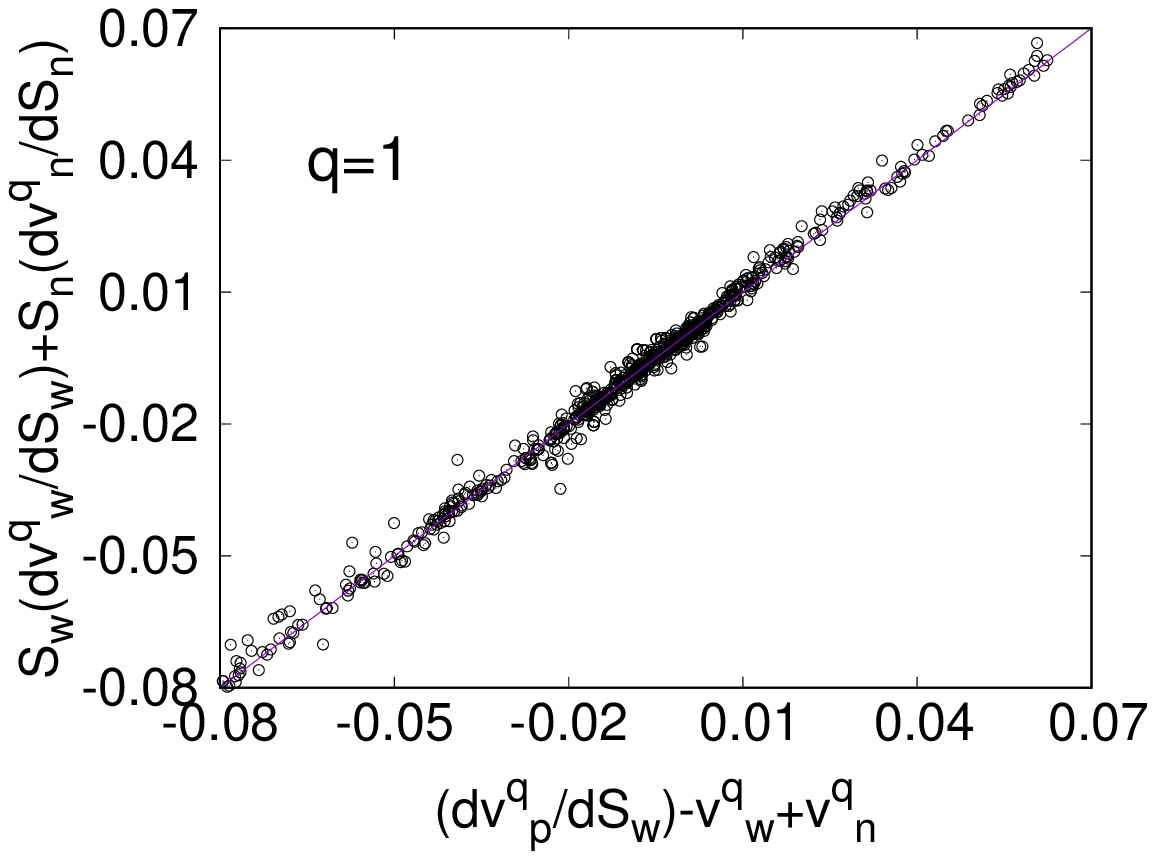}
  \includegraphics[width=0.3\textwidth,clip]{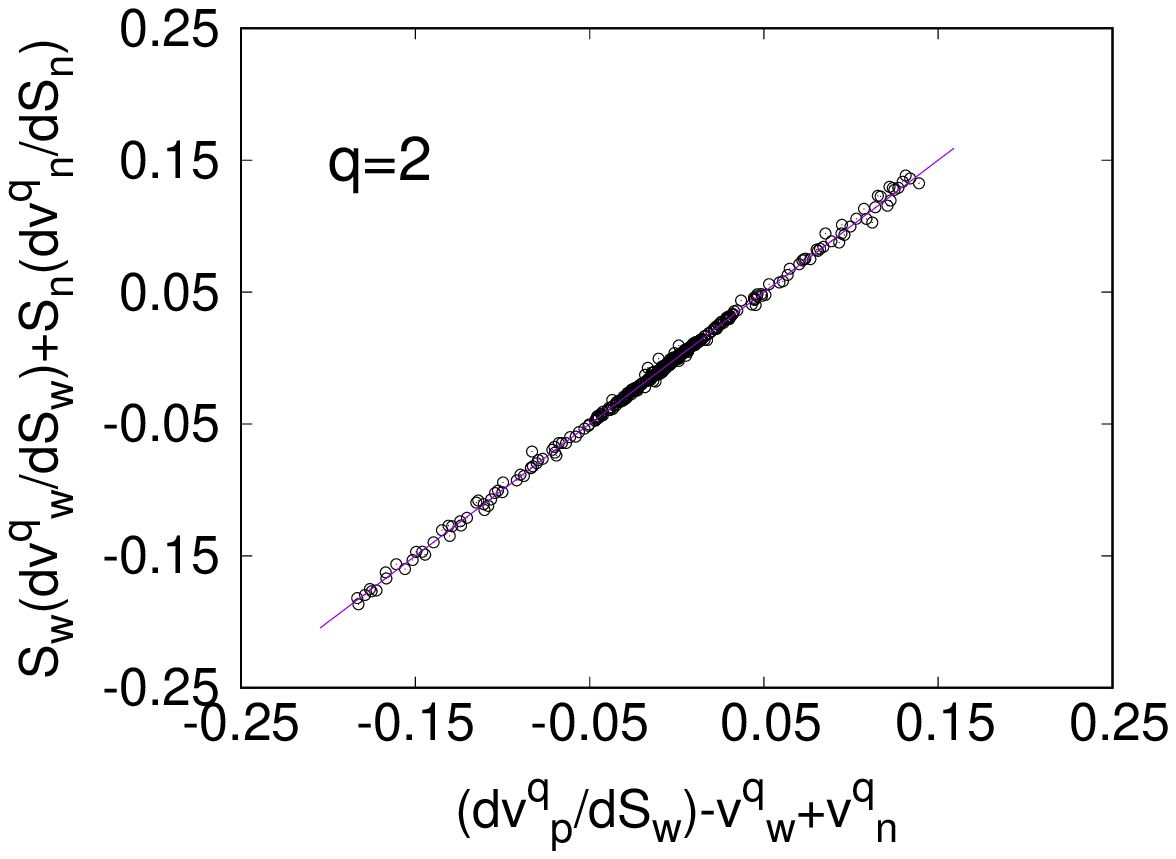}
  \includegraphics[width=0.3\textwidth,clip]{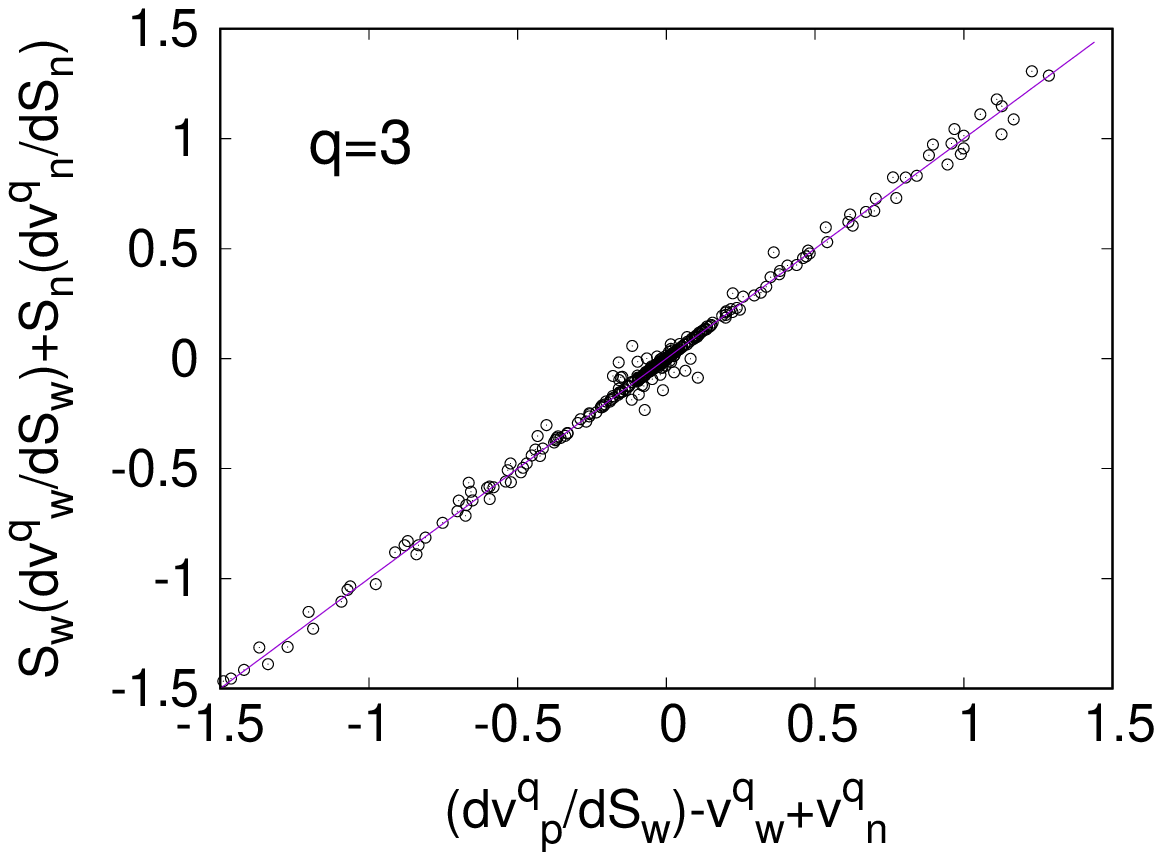}
  \includegraphics[width=0.3\textwidth,clip]{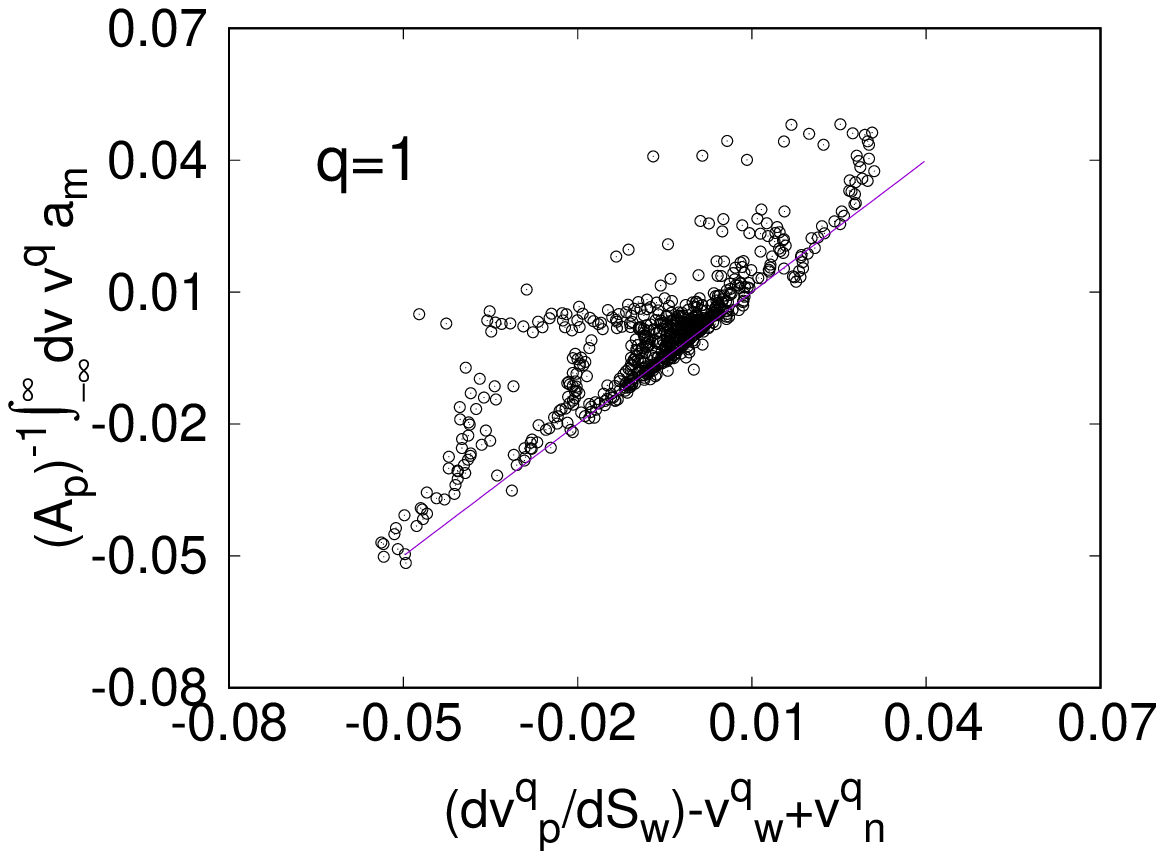}
  \includegraphics[width=0.3\textwidth,clip]{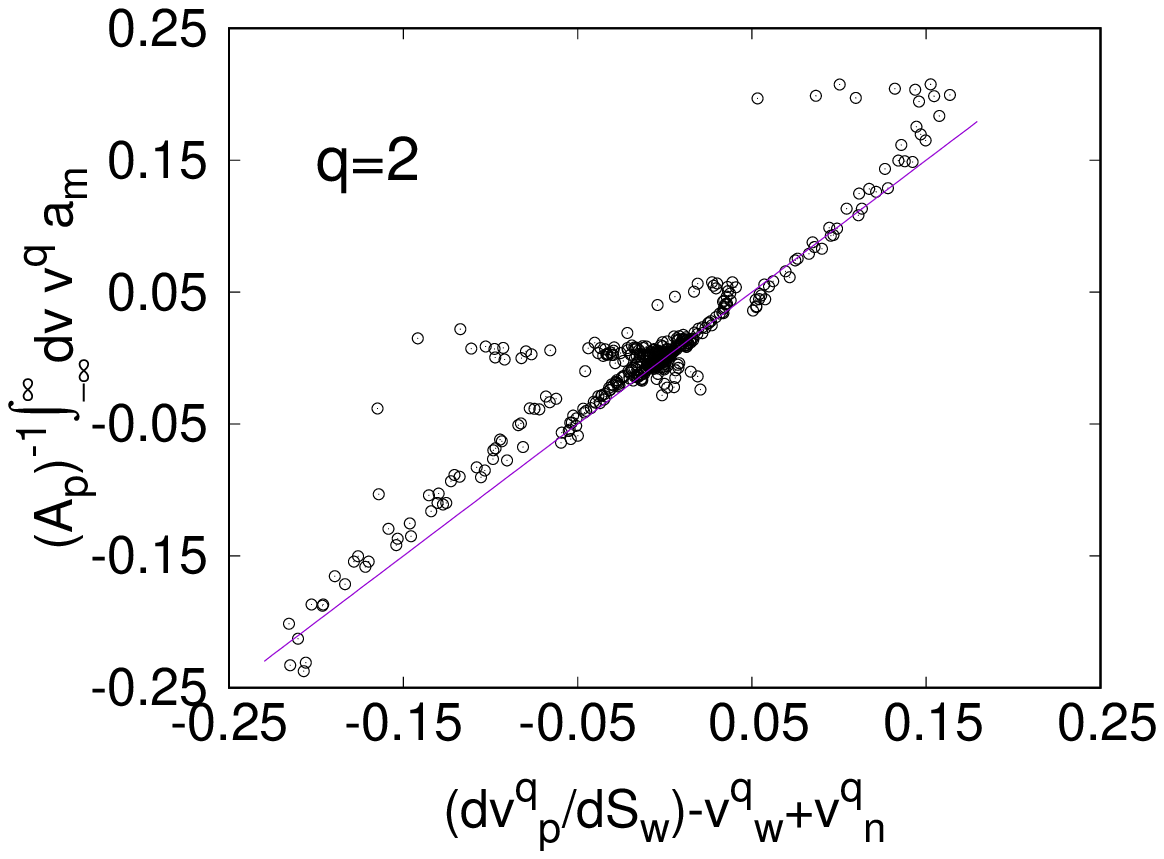}
  \includegraphics[width=0.3\textwidth,clip]{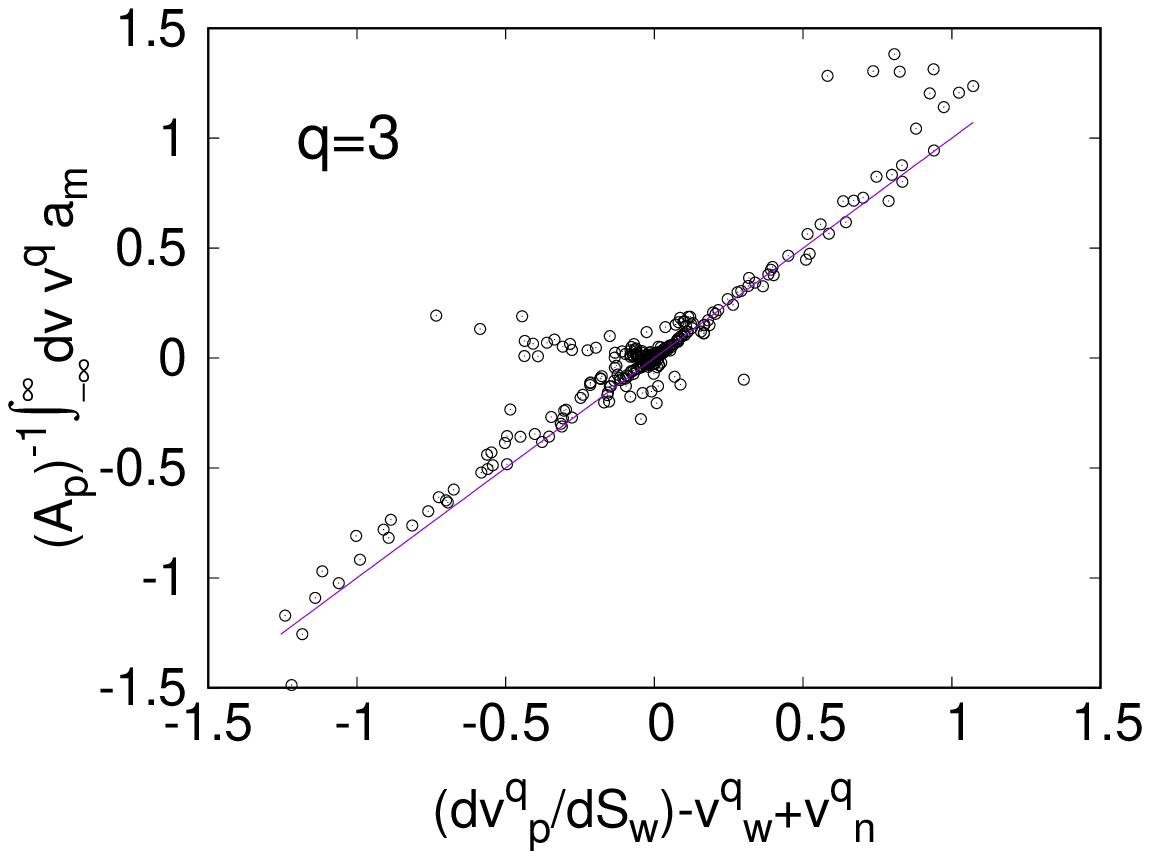} 
\caption{Measurements of $v_m^q$ for the 3D Berea network where the
  top row corresponds to the direct measurements using equations
  (\ref{eqn18-1}) and (\ref{eqn20-1}), and the bottom row corresponds
  to the measurement from the differential area distributions using
  equation (\ref{eqn3.8}). Here, larger fluctuations in the results
  calculated with the differential pore area are observed compared to
  the 2D network.}
\label{fig5}
\end{figure*}
%--------------------------------------------------------------------
For the second approach, we construct the distribution of differential
transversal pore areas $a_p$, $a_w$ and $a_n$ such that $a_pdv$, $a_wdv$ and
$a_ndv$ express the transversal pore areas for the total, wetting and non-wetting
fluids within the velocity range from $v$ to $v+dv$, so that they
satisfy equation (\ref{eqn3.1}), (\ref{eqn3.10}) and
(\ref{eqn3.3}). We therefore have,
\begin{align}
\label{eqn5.06}
a_p(v)dv & = \displaystyle\frac{1}{L}\displaystyle\sum_ja_jl_{x,j}\;, \nonumber \\ 
a_w(v)dv & = \displaystyle\frac{1}{L}\displaystyle\sum_ja_jl_{x,j}S_{w,j}\;, \nonumber \\
a_n(v)dv & = \displaystyle\frac{1}{L}\displaystyle\sum_ja_jl_{x,j}S_{n,j}\;,
\end{align}
where $j$ runs over all the sites satisfying the condition:
$v<v_j<v+dv$, $v_j$ being the local velocity of link $j$. In case of
the 2D lattice, $l_{x,j}$s are same for any $j$ and given by
$l_{x,j}=l/\sqrt{2}$. With these, different moments of the seepage
velocities are then calculated using equations (\ref{eqn4.1}),
(\ref{eqn4.3}) and (\ref{eqn4.4}) respectively.

For any saturation, the seepage velocities and their higher moments
should follow the relations (\ref{eqn5.4}), (\ref{eqn3.7}) and
(\ref{eqn4.2}). We plot our numerical measurements in figure
\ref{fig1} and \ref{fig2} for 2D and 3D respectively. The upper row in
each figure corresponds to the direct measurements and the lower row
correspond to the measurements from the differential area
distribution. A good agreement with the relations can be observed for
first as well as for the higher moments for both the networks.

Next we measure the fluctuations in the seepage velocities which obey
equation (\ref{eqn4.13}). Numerically, $\Delta v_p^2$, $\Delta v_w^2$
and $\Delta v_n^2$ are calculated from the knowledge of the $1^{\rm
  st}$ and $2^{\rm nd}$ moments by,
\begin{align}
\label{eqn5.06}
\Delta v_p^2 & = \langle v^2 \rangle_p - \langle v \rangle^2_p \;, \nonumber \\
\Delta v_w^2 & = \langle v^2 \rangle_w - \langle v \rangle^2_w \;, \nonumber \\
\Delta v_n^2 & = \langle v^2 \rangle_n - \langle v \rangle^2_n \;.
\end{align}
In figure \ref{fig3}, we plot these fluctuations for the two networks
to compare with equation (\ref{eqn4.13}) and good agreements is
observed. There are some deviations in the results for the Berea
network, since the results in 3D is based on only one network
configuration whereas the results for 2D are averaged over $10$
different configurations.

Finally, we verify the relations between seepage velocities and their
higher moments while varying the fluid saturation as given by the
equations (\ref{eqn18-1}), (\ref{eqn20-1}) and (\ref{eqn4.22}). For
this, we first calculated the co-moving velocity ($v_m$) and its
higher moments from equations (\ref{eqn18-1}) and (\ref{eqn20-1})
where we used the values of the seepage velocities measured with the
direct approach. This is shown in the top rows of figures \ref{fig4}
and \ref{fig5} for 2D and 3D respectively which show good agreements
with equations (\ref{eqn18-1}) and (\ref{eqn20-1}). We then compare
these values of $v_m^q$ with the measurements from the differential
transversal areas using equation (\ref{eqn3.8}). For this, we first
constructed the histogram for the differential pore area $a_m$
corresponding to the co-moving velocity from equation (\ref{eqn3.9})
where we have used the variations of $a_p$, $a_w$ and $a_n$ with the
saturation $S_w$. For this purpose, we have considered $21$ different
values of saturations within $0$ and $1$ with an interval of
$0.05$. We then integrate $a_m$ from $-\infty$ to $\infty$, weighted
by the velocity and normalized by the total pore area to obtain the
desired co-moving velocity with equation (\ref{eqn3.8}). These results
are plotted in the bottom row of figures \ref{fig4} and \ref{fig5}
where they are compared with the results from direct measurements. The
data points roughly follow the diagonal straight line showing
satisfactory agreement with the theoretical formulations. However, we
observe deviations in the results that is higher compared to the
direct measurements. We believe this is due to the numerical errors
that added up from several steps in the calculation such as the
binning techniques while measuring the distributions, taking the
derivatives and calculating the integrals. Moreover, the fluctuations
for 3D are much higher compared to 2D, which is due to the lack of
averaging over different samples as we have already mentioned earlier.

%--------------------------------------------------------------------
\section{Summary}
\label{secCon}

The aim of this paper is to provide the link between the pseudo-thermodynamic
theory at the continuum level developed in \cite{hsbkgv18} (see Section \ref{secThermo})
and the velocities occurring at the pore level during immiscible two-phase flow in porous 
media.  This link is provided by defining the differential transversal pore areas
defined in Section \ref{secDiff},
which essentially correspond to the statistical distributions of velocities at
the pore level.  The central quantities are the velocity differential transversal
pore area $a_p$, the wetting fluid differential velocity transversal
pore area $a_w$, the non-wetting fluid velocity differential transversal
pore area $a_n$, and the co-moving velocity differential transversal
pore area $a_m$.  We also consider the thermodynamic velocity differential transversal
pore areas $\hat{a}_w$ and $\hat{a}_n$.  The relations found by
Hansen et al.\ \cite{hsbkgv18} for the average seepage velocities, the 
co-moving velocity and the thermodynamic velocities are generalized to the 
differential transversal areas here.  In the following Section \ref{secMoments}
the relations are generalized to higher moments of the velocity distributions. 
     
The theoretical derivations are then in Section \ref{secNum} validated by numerical
simulations. We used dynamic pore-network modeling where an
interface-tracking model is used to simulate steady-state two-phase
flow.  We used both regular pore networks and an irregular
pore network reconstructed from a Berea sandstone for our simulations. 
By measuring the seepage velocities from the differential area distributions and 
comparing them with the direct measurements, we validated the essential 
predictions from the earlier theoretical sections. 

%--------------------------------------------------------------------
\section*{Author Contributions}

SR did the numerical simulations for 2D and data analysis, SS developed the codes and performed 3D simulations.
AH developed the theory.  The paper was written by all three authors contributing equally.

%--------------------------------------------------------------------
\section*{Acknowledgment}

The authors thank Carl Fredrik Berg, M. Aa. Gjennestad, Knut J{\o}rgen
M{\aa}l{\o}y, Per Arne Slotte, Ole Tors{\ae}ter, Morten Vassvik and
Mathias Winkler for interesting discussions. AH thanks the Beijing
Computational Science Research Center CSRC for the hospitality and
financial support. SS was supported by the National Natural Science
Foundation of China under grant number 11750110430. This work was
partly supported by the Research Council of Norway through its Center
of Excellence funding scheme, project number 262644.
%--------------------------------------------------------------------

%--------------------------------------------------------------------
\end{document}